\documentclass{aa}  
\usepackage{graphicx}
\usepackage{txfonts}
\usepackage{hyperref}
\begin{document}

   \title{A super-Earth on a close-in orbit around the M1V star GJ 740
   \thanks{Based on observations made with the Italian Telescopio Nazionale Galileo (TNG), operated on the island of La Palma by the INAF - Fundación Galileo Galilei at the Roche de Los Muchachos Observatory of the Instituto de Astrofísica de Canarias (IAC); and the CARMENES instrument installed at the 3.5m telescope of the Calar Alto Observatory, Spain.}
   \thanks{The RVs used in this paper are available in electronic form at the CDS via anonymous ftp to \url{cdsarc.u-strasbg.fr} (130.79.128.5) or via \url{http://cdsweb.u-strasbg.fr/cgi-bin/qcat?J/A+A/}}
   }

   \subtitle{A HADES and CARMENES collaboration}

   \author{B. Toledo-Padr\'on\inst{1,2} 
                  A. Su\'arez Mascare\~no\inst{1,2}, 
                  J. I. Gonz\'alez Hern\'andez\inst{1,2}, 
                  R. Rebolo\inst{1,2,3}, 
                  M. Pinamonti\inst{4},\newline 
                  M. Perger\inst{5,6}, 
                  G. Scandariato\inst{7}, 
                  M. Damasso\inst{4}, 
                  A. Sozzetti\inst{4}, 
                  J. Maldonado\inst{8}, 
                  S. Desidera\inst{9}, 
                  I. Ribas\inst{5,6}, 
                  G. Micela\inst{8},\newline 
                  L. Affer\inst{8}, 
                  E. Gonz\'alez-Alvarez\inst{10}, 
                  G. Leto\inst{7}, 
                  I. Pagano\inst{7}, 
                  R. Zanmar Sánchez\inst{7}, 
                  P. Giacobbe\inst{4}, 
                  E. Herrero\inst{5,6},\newline 
                  J. C. Morales\inst{5,6}, 
                  P. J. Amado\inst{11},
          J. A. Caballero\inst{12},
          A. Quirrenbach\inst{13},
          A. Reiners\inst{14},
          \and
          M. Zechmeister\inst{14}
          }
          
   \titlerunning{A super-Earth on a close-in orbit around the M1V star GJ 740}
          
   \authorrunning{B. Toledo-Padr\'on et al.}

   \institute{Instituto de Astrof\'isica de Canarias, E-38205 La Laguna, Tenerife, Spain \\
              \email{btoledo@iac.es}
              \and
                          Universidad de La Laguna, Departamento de Astrof\'isica, E-38206 La Laguna, Tenerife, Spain
                          \and
                          Consejo Superior de Investigaciones Cient\'ificas, E-28006 Madrid, Spain
                          \and
                          INAF-Osservatorio Astrofisico di Torino, via Osservatorio 20, 10025 Pino Torinese, Italia
                          \and
                          Institut de Ci\`encies de l'Espai, Campus UAB, C/Can Magrans s/n, 08193 Bellaterra, Spain
                          \and
                          Institut d'Estudis Espacials de Catalunya (IEEC), 08034 Barcelona, Spain
                          \and
                          INAF-Osservatorio Astrofisico di Catania, via S. Sofia 78, 95123 Catania, Italia
                          \and
                          INAF-Osservatorio Astronomico di Palermo, Piazza Parlamento 1, 90134 Palermo, Italy
                          \and
                          INAF-Osservatorio Astronomico di Padova, Vicolo dell’Osservatorio 5, 35122 Padova, Italia
                          \and
                          Centro de Astrobiología (CSIC-INTA), Carretera de Ajalvir km 4, 28850 Torrejón de Ardoz, Madrid, Spain
                          \and
                          Instituto de Astrofísica de Andalucía (IAA-CSIC), Glorieta de la Astronomía s/n, 18008 Granada, Spain
                          \and
                          Centro de Astrobiología (CSIC–INTA), ESAC, Camino Bajo del Castillo s/n, 28691 Villanueva de la Cañada, Madrid, Spain
                          \and
                          Landessternwarte, Zentrum für Astronomie der Universität Heidelberg, Königstuhl 12, 69117 Heidelberg, Germany
                          \and
                          Institut für Astrophysik, Georg-August-Universität, Friedrich-Hund-Platz 1, 37077 Göttingen, Germany
             }

   \date{\centering Received December 10, 2020 / Accepted February 14, 2021}

 
  \abstract
   {M-dwarfs have proven to be ideal targets for planetary radial velocity (RV) searches due to their higher planet-star mass contrast, which favors the detection of low-mass planets. The abundance of super-Earth and Earth-like planets detected around this type of star motivates further such research on hosts without reported planetary companions.}
   {The HADES and CARMENES programs are aimed at carrying out extensive searches of exoplanetary systems around M-type stars in the northern hemisphere, allowing us to address, in a statistical sense, the properties of the planets orbiting these objects. In this work, we perform a spectroscopic and photometric study of one of the program stars (GJ~740), which exhibits a short-period RV signal that is compatible with a planetary companion.} 
   {We carried out a spectroscopic analysis based on 129 HARPS-N spectra taken over a time span of 6\,yr combined with 57 HARPS spectra taken over 4\,yr, as well as 32 CARMENES spectra taken during more than 1\,yr, resulting in a dataset with a time coverage of 10\,yr. We also relied on 459 measurements from the public ASAS survey with a time-coverage of 8\,yr, along with 5\,yr of photometric magnitudes from the EXORAP project taken in the $V$, $B$, $R,$ and $I$ filters to carry out a photometric study. Both analyses were made using Markov Chain Monte Carlo (MCMC) simulations and Gaussian Process regression to model the activity of the star.}
   {We present the discovery of a short-period super-Earth with an orbital period of 2.37756$^{+0.00013}_{-0.00011}$\,d and a minimum mass of 2.96$^{+0.50}_{-0.48}$\,M$_{\oplus}$. We offer an update to the previously reported characterization of the magnetic cycle and rotation period of the star, obtaining values of $P_{\rm rot}$=35.563~$\pm$0.071\,d and $P_{\rm cycle}$=2800$\pm$150\,d. Furthermore, the RV time series exhibits a possibly periodic long-term signal, which might be related to a Saturn-mass planet of~$\sim$~100\,M$_{\oplus}$.}
   {}
   

   \maketitle

\section{Introduction}

The development of second and third-generation echelle spectrographs has produced a significant boost in the radial velocity (RV) searches for Earth-like planets around M-dwarfs. Since the first discovery of an exoplanet orbiting an M-type star using the RV method \citep{1998A&A...338L..67D,1998ApJ...505L.147M}, a total of 116 planets have been discovered around M-dwarfs using this technique \footnote{\url{https:/exoplanets.nasa.gov/}}. The vast majority of them (75\%) have been detected in the last decade via such instruments as HARPS-N (e.g., \citealt{2019A&A...622A.193A,2019A&A...625A.126P}), HARPS (e.g., \citealt{2020MNRAS.493..536D,2020A&A...633A..44G}), CARMENES (e.g., \citealt{2019A&A...627A.116L,2019A&A...627A..49Z}), SOPHIE (e.g., \citealt{2019A&A...625A..18H}), and PFS (e.g., \citealt{2019ApJS..242...25F}).

Despite the fact that M-dwarfs are the most common stars in the Milky Way \citep{2000ARA&A..38..337C,2015AJ....149....5W}, only 10\% of all known planetary companions have been detected around this type of star. From the complete four-year \textit{Kepler} dataset, \citet{2015ApJ...807...45D} found 156 planet candidates orbiting M-type stars, estimating an average of~$\sim$~2.5 planets per star in the $P_{\rm orb}<$200\,d and $R$=1--4\,R$_{\rm \oplus}$ regime. This occurrence rate, combined with their closer habitable zones due to their lower luminosities, makes such low-mass stars ideal targets for the search of temperate Earth-like planets. However, the complexity of the characteristic stellar activity pattern of these stars requires a careful analysis of the chromospheric activity indicators to identify false planetary signals induced by the rotation of the star \citep{2007A&A...474..293B,2014Sci...345..440R,2018A&A...612A..89S,2019MNRAS.488.5145T}.

The planetary formation scenario provided by the core accretion theory \citep{1996Icar..124...62P} indicates that the most common planets around M-dwarfs are super-Earth and Neptune-mass planets \citep{2018A&A...609L...5R,2018A&A...620A.171L,2019A&A...624A.123P,2020A&A...637A..93G}, most of them located in the $P_{\rm orb}<$100\,d region. Nonetheless, there have also been some detections of gas giants orbiting late-type stars \citep{2006PASP..118.1685B,2008Sci...319..927G,2010ApJ...721.1467H,2019Sci...365.1441M}, although their location tends to be beyond the snow line of the system (where volatile compounds such as water or carbon dioxide could condense into solid ice grains). \citet{2006ApJ...649..436E} estimated a frequency $<$ 1.27 \% of Jovian planets around M-dwarfs, which is lower than that of FKG-type stars. The elusiveness of giant planets orbiting M-type stars is contrasted by the higher occurrence rate of super-Earth planets \citep{2014MNRAS.441.1545T}, with some Earth-like planets also being reported \citep{2016Natur.536..437A,2017A&A...605L..11A,2018A&A...613A..25B,2019A&A...627A..49Z,2020MNRAS.493..536D}, but still lacking in tems of proper statistics of their population and occurrence rate at present.

We report the discovery of a short-period super-Earth orbiting the nearby M-type star GJ~740, as part of the HADES (HArps-n red Dwarf Exoplanet Survey) program, complemented by data from HARPS and the CARMENES survey. HADES is a collaboration between the Instituto de Astrofísica de Canarias (IAC), the Institut de Ciències de l’Espai (IEEC-CSIC), and the Italian GAPS (Global Architecture of Planetary Systems) program \citep{2013A&A...554A..28C} and it is aimed at detecting and characterizing exoplanets around M-dwarfs, as well as improving the available statistical information on the properties of these objects. The reported detection contributes to the planetary discoveries made by the collaboration \citep{2016A&A...593A.117A,2017A&A...608A..63P,2017A&A...605A..92S,2018A&A...617A.104P,2019A&A...622A.193A,2019A&A...624A.123P,2019A&A...625A.126P}. In this work, we also explore the possibility of an outer high-mass planetary companion characterized by a long orbital period and its impact on the planetary configuration.

The paper is structured as follows. Section~\ref{sec:Data} presents the dataset used in this work, including both spectroscopy and photometry. Section~\ref{sec:GJ740} details the stellar properties of GJ~740. Section~\ref{sec:Method} demonstrates the techniques used for the RV and stellar activity indicator measurements. Section~ \ref{sec:Analysis} describes the analysis of these and the photometric measurements. Section~\ref{sec:Discussion} features a discussion of these results and Sect.~\ref{sec:Conclusions} provides our conclusions.

\section{Data}

\label{sec:Data}

\subsection{Spectroscopic dataset}

The HADES RV program monitored GJ~740 from 26 May 2013 (BJD=2456438.7) to 16 June 2019 (BJD=2458650.7) using the HARPS-N spectrograph \citep{2012SPIE.8446E..1VC} installed at the 3.6m Telescopio Nazionale Galileo (TNG) in the Roque de los Muchachos Observatory, Spain. The high resolution of~R~$\sim$~115\,000 provided by this fiber-fed echelle spectrograph, along with its spectral range from 3830 to 6900\,\AA \ are ideal for high-precision RV searches. We obtained a total of 129 HARPS-N spectra over a time span of 6\,yr. Every spectrum was taken using an exposure time of 900\,s to average out the short-time periodic oscillations of the star \citep{2011IAUS..276..527D}, although this phenomenon has not yet been detected in M-dwarfs \citep{2017MNRAS.469.4268B}. The average signal-to-noise ratio (S/N) achieved at 5500\,\AA \ per pixel was 110, enough to ensure a good exposure level of the blue part of the spectra, which contains the Ca~\textsc{II} H\&K lines that are especially weak for M-type stars \citep{1989ApJ...345..536G,2011arXiv1107.5325L}. Some of the spectra were obtained using a Fabry-Pérot (FP) interferometer \citep{2010SPIE.7735E..4XW} for the wavelength calibration due to the lack of availability of the Th-Ar lamp for the simultaneous calibration. In those cases, the Th-Ar lamp was used to obtain the absolute calibration prior to the observations, and then all the FP spectra were referred to the Th-Ar calibration spectrum with a drift value. The interferometer provides a high level of short-term RV precision and has the advantage of avoiding possible contamination of the Ca~\textsc{II} H\&K lines, but lacks the precise drift correction provided by the Th-Ar lamp. However, the mean inter-night instrumental drift calculated by \citet{2017A&A...598A..26P} for the whole HADES sample was about 1\,m\,s$^{-1}$.

We also acquired 32 spectra with the CARMENES spectrograph at the Calar Alto Observatory \citep{2018SPIE10702E..0WQ} overlapping at the epoch during which the HARPS-N observations were carried out. CARMENES allows for the simultaneous observation in two different channels that cover the visible (between 5200 and 9600\,\AA) and near-infrared (between 9600 and 17\,100\,\AA) regions of the spectra, with resolutions of~R~$\sim$~94\,600 and R~$\sim$~80\,400, respectively. The RV precision provided by both channels is~$\sim$~1\,m\,s$^{-1}$, similar to the one obtained in HARPS-N. The CARMENES spectra were acquired from 11th April 2016 (BJD=2457489.7) to 27th July 2017 (BJD=2457962.5). This dataset is characterized by an average S/N per pixel of 122. Hollow cathode lamps combined with an FP etalon were used to obtain the wavelength calibration of these spectra \citep{2015A&A...581A.117B}. 

Additionally, this star has been monitored from the southern hemisphere using the HARPS spectrograph \citep{2003Msngr.114...20M} installed at the 3.6m telescope of La Silla Observatory, Chile. This fiber-fed echelle spectrograph has similar characteristics to its northern counterpart. It is contained in a vacuum vessel to minimize the RV drifts produced by temperature and pressure variations. We used 57 HARPS spectra from the ESO public database taken over a time span of more than 4\,yr: from 30 June 2008 (BJD=2454647.7) to 10 August 2012 (BJD=2456149.7). The exposure time used was the same as for the HARPS-N spectra, achieving an average S/N per pixel at 5500 \AA \ of 81.5. In this case, the majority of the spectra were calibrated with an FP interferometer.

\subsection{Photometric dataset}

To complement the spectroscopic analysis, we used 474 photometric measurements from ASAS (All Sky Automated Survey) \citep{1997AcA....47..467P}. The measurements come from one of the three observing stations of the project (ASAS-S/ASAS-3), located at the Las Campanas Observatory, Chile. This survey carries out observations  in $V$ and $I$ bands simultaneously, with a plate scale of 14"/pixel and an average accuracy of~$\sim$~0.05\,mag per exposure, achieving the best results in the range of $V$~$\sim$~8--12\,mag (which includes GJ~740). The light curve provided by ASAS includes a quality flag associated with every measurement. Using these flags, we are able to discard the worst-quality data (labeled as grade `C' and `D'). This leaves us with 459 measurements covering a time span of eight years: from 26 September 2001 (BJD=2452178.6) to 2 November 2009 (BJD=2455137.5). Since the original data were given in the Heliocentric Julian Date (HJD), we performed a conversion to the Barycentric Julian Date (BJD) considering the difference between the two systems of up to 8\,s due to the motion of the Sun (the reference frame in this case), caused mainly by Jupiter and Saturn \citep{2010PASP..122..935E}.

We also relied on the publicly available light curve from the SuperWASP database \citep{2006PASP..118.1407P}. The data were taken with the survey facilities located at the South African Astronomical Observatory and the Roque de los Muchachos Observatory. The cameras used feature a plate scale of 13.7"/pixel and an average accuracy better than 1\% for objects with $V$~$\sim$~7.0--11.5. We analyzed 2350 photometric data points taken between 18 June 2006 (BJD=2453904.7) and 23 July 2008 (BJD=2454670.7). A time conversion from HJD to BJD was performed based on the procedure used for the ASAS dataset.

Within the framework of the EXORAP project, we collected~$\sim$5\,yr of $B$, $V$, $R$, and $I$-band photometry at the INAF-Catania Astrophysical Observatory with an 80 cm f/8 Ritchey-Chretien robotic telescope (APT2) located at Serra la Nave on Mt. Etna. We performed data reduction by overscan, bias, dark subtraction, and flat fielding with IRAF\footnote{IRAF is distributed by the National Optical Astronomy Observatories, which are operated by the Association of Universities for Research in Astronomy, Inc., under a cooperative agreement with the National Science Foundation.} procedures and we visually inspected the data to check the quality (see \citealp{2016A&A...593A.117A} for details). Errors in the individual photometric points include the intrinsic noise (photon noise and sky noise) and the root-mean-square (RMS) of the ensemble stars used for computing the differential photometry. The final dataset contains~$\sim$~200 photometric points for each of the $B$, $V$, and $R$ bands distributed over five consecutive seasons, between BJD=2456480.5 and BJD=2458003.5. Due to technical reasons, the $I$-band photometry contains only 100 points covering the first four seasons.

\section{GJ 740}

\label{sec:GJ740}

GJ~740 (HD~176029, BD+05~3993) is a bright ($m_{\rm V}$=9.2) M1V-type high-proper motion star, located at 11.1 pc from the Sun \citep{2018A&A...616A...1G}. The search for companions of this star has produced null detections \citep{2005AJ....130.1212C,2020AJ....159..139L}, reporting only some faint field background objects \citep{2005AJ....130.1212C}. Stellar parameters from the literature are shown in Table~\ref{tab:GJ740_Properties} along with new values established in this work.

\renewcommand{\arraystretch}{1.25}

\begin{table}
\centering
        \caption{Stellar properties of GJ~740.}
        \label{tab:GJ740_Properties}
        \begin{tabular}{lccr}
                \hline
                Parameter & GJ 740 & Ref.\\
                \hline
                $\mathrm{RA}$ (J2000)                        & 18:58:00.14             & [1] \\
                $\mathrm{Dec}$ (J2000)                       & +05:54:29.24            & [1] \\
                $\mu_{\alpha}$ cos $\delta$ [mas\,yr$^{-1}$] & $-$196.301 $\pm$ 0.087  & [1] \\
                $\mu_{\delta}$ [mas\,yr$^{-1}$]              & $-$1220.467 $\pm$ 0.092 & [1] \\
                Distance [pc]                                & 11.1017 $\pm$ 0.0061    & [2] \\
                $m_{\rm B}$                                  & 10.639                  & [3] \\
                $m_{\rm V}$                                  & 9.367                   & [3] \\
                Spectral type                                & M1.0V                   & [4] \\
                \textit{T}$_{\rm eff}$ [K]                   & 3913 $\pm$ 51           & [5] \\
                $\log g$ [cgs]                               & 4.68 $\pm$ 0.07         & [5] \\
                $[\mathrm{Fe/H}]$ [dex]                      & 0.08 $\pm$ 0.16         & [5] \\
                $M_{\star}$ [M$_{\odot}$]                    & 0.58 $\pm$ 0.06         & [4] \\
                $R_{\star}$ [R$_{\odot}$]                    & 0.56 $\pm$ 0.06         & [4] \\
                $\log (L_{\star}$/L$_{\odot}$)               & $-$1.206 $\pm$ 0.097    & [4] \\
                $\log (L_{\rm x}$/L$_{\rm bol}$)             & $-$4.85  $\pm$ 0.17     & [6] \\
                $v \sin i$ [km\,s$^{-1}$]                    & 0.92  $\pm$ 0.59        & [4] \\
                $i$ [deg]                                    & $>$ 25                  & [7] \\
                $a_{\rm sec}$ [m\,s$^{-1}$\,yr$^{-1}$]       & 0.3903 $\pm$ 0.0005     & [8] \\
                $\log_{10}$ ($R^{'}_{\rm HK}$)               & $-$4.88 $\pm$ 0.05      & [9] \\
                $P_{\rm rot}$ [d]                            & 35.563~$\pm$0.071       & [9] \\  
                \hline
        \end{tabular}
        \begin{minipage}{\columnwidth}
{\footnotesize \textbf{References:} [1] \citet{2018A&A...616A...1G}; [2] \citet{2018AJ....156...58B} [3] \citet{2020A&A...642A.115C}; [4] \citet{2017A&A...598A..27M}; [5] \citet{2018A&A...615A...6P}; [6] \citet{2019A&A...624A..27G}; [7] \citet{2018A&A...612A..89S} ; [8] Calculated following \citet{2009A&A...505..859Z}; [9] This work.}
\end{minipage}  
\end{table}

This star has been studied by \citet{2017A&A...600A..13A}, who obtained a rotation period of 24\,d using 55 HARPS spectra. GJ~740 was included within a sample of 107 M dwarfs studied by \citet{2020MNRAS.491.5216G} as part of the APACHE  photometric transit search project \citep{2013EPJWC..4703006S}, recovering a period of 35.6422$\pm$0.0063\,d from a 765\,d time span dataset. Through another photometric study based on the ASAS survey, \citet{2019A&A...621A.126D} obtained a rotation period of 35.20$\pm$0.30\,d. \citet{2018A&A...612A..89S} carried out a stellar activity analysis on the whole HADES sample, deriving an expected rotational period of $P_{\rm rot}$=36.3~$\pm$~1.7\,d for this star adding 93 HARPS-N to the existing HARPS spectra, along with 458 ASAS photometric measurements. This study provided also a mean level of chromospheric activity of $\log(R^{'}_{\rm HK})$=$-$4.88~$\pm$~0.04 computed through the Ca~\textsc{II} H\&K lines, along with a magnetic cycle of 2044\,d detected in several stellar activity indicators. We extend this stellar activity study with new HARPS-N and CARMENES spectra in order to model the chromospheric effects on the RV measurements.

\section{Determination of RVs and stellar activity indicators}

\label{sec:Method}

For the RV calculation in the HARPS-N and HARPS datasets, we followed two different approaches. The first one is based on the HARPS-N DRS (Data Reduction Software) pipeline \citep{2007A&A...468.1115L}, which builds a cross-correlation function (CCF) for a certain template mask driven by the spectral type. For GJ~740, we used the M2 mask, which contains 9196 wavelength intervals of 0.02 \AA \ width with different depths located at the positions of isolated stellar lines. This mask is shifted 161 times in an RV range of~$\sim$~40\,km\,s$^{-1}$, providing a CCF for each echelle order. The resulting CCFs are fitted by a Gaussian along with an offset constant to obtain an RV measurement per echelle order. These measurements are weighted in terms of the flux of each order to obtain the final RV value. We modify this last step of the process by weighting each CCF order according to its mean flux, performing a secular acceleration correction based on the proper motion of the star, and fitting the combined CCF with a Gaussian along with a second-order polynomial, which provides slightly better results. The RVs computed through this methodology exhibit an RMS=4.3\,m\,s$^{-1}$ with a mean RV uncertainty of 1.4\,m\,s$^{-1}$ in the case of HARPS and an RMS=6.3\,m\,s$^{-1}$ with a mean RV uncertainty error of 1.2\,m\,s$^{-1}$ in the case of HARPS-N. The difference between both datasets is caused by the RV trend present in the HARPS-N measurements.

The second approach that we implemented for computing the RVs of the HARPS-N and HARPS spectra is based on the TERRA  reduction software \citep{2012ApJS..200...15A}, which corrects the spectra for the blaze function, secular acceleration \citep{2003A&A...403.1077K}, and then puts the spectra in the solar system barycentric frame by correcting for the barycentric and stellar RV. Then it performs a template matching with the highest S/N spectra, carrying out a least-squares fit on the residuals for every shift and providing an RV value from the minimum $\chi^{2}$ value. The RVs computed by TERRA yield an RMS=4.3\,m\,s$^{-1}$ and a mean uncertainty error of 0.9\,m\,s$^{-1}$ in the case of HARPS and RMS=5.8\,m\,s$^{-1}$ and mean uncertainty error of 0.8\,m\,s$^{-1}$ in the case of HARPS-N. These results support the superiority of performance on the part of the TERRA pipeline for M-dwarfs \citep{2017A&A...598A..26P} and for this reason, we opted for this reduction procedure in our analysis.

In the case of the CARMENES dataset, the spectra were reduced using the CARACAL pipeline \citep{2016SPIE.9910E..0EC} and the RVs were computed using the SERVAL tool \citep{2018A&A...609A..12Z}, which is a public code \footnote{\url{https://github.com/mzechmeister/serval}} similar to TERRA. This tool performs a template-matching using a high S/N template built from the available spectra of the star as a reference. The CARMENES RV time series computed by SERVAL presents an RMS=4.3\,m\,s$^{-1}$, with a mean uncertainty of 1.8\,m\,s$^{-1}$.

In order to model the stellar activity effects on the RV measurements, we studied the flux variations of certain spectral lines connected with several chromospheric activity indicators. The measurements were made on the spectra corrected from the blaze function (using a blaze spectrum given by the DRS pipeline), the barycentric velocity of the Earth, and the radial velocity of the star. We also carried out a re-binning of the spectra in order to obtain the same wavelength width per pixel (0.01 \AA). To avoid any wavelength shifts that can affect the flux measurements, we correlated the spectra using the first spectrum taken by each spectrograph as a reference. All these spectra were co-added into an average spectrum that was normalized and used for calculating the weight of each echelle order for the individual spectra. Once all these corrections had been applied to the individual spectra, we derived three activity indicators: H${\rm \alpha}$ \citep{2003A&A...403.1077K,2011A&A...534A..30G}, the S$_{\rm mw}$ index associated with the Ca~\textsc{II} H\&K lines \citep{1984ApJ...279..763N,2011arXiv1107.5325L}, and the NaD index associated with the doublet of NaI D$_{1}$ and D$_{2}$ lines \citep{2007MNRAS.378.1007D,2009MNRAS.400..238H}. The continuum passbands of this last index were modified following \citet{2019MNRAS.488.5145T} in order to have them in the same echelle order of the core lines, avoiding the overlap zones between spectral orders. Moreover, we calibrated the Mount-Wilson S-index following \citet{2011arXiv1107.5325L} and \citet{2015MNRAS.452.2745S}. The CARMENES spectra do not cover this index in their wavelength range. The error on all the measurements was calculated using the RMS on the spectral bands related to each index (core lines and continuum passbands), along with error propagation.

The stellar activity can also be tracked using properties related to the CCF such as the full width at half maximum (FWHM) or the bisector span (BIS). We measured the FWHM using the CCFs from the DRS pipeline weighted, corrected for secular acceleration, and fitted by a Gaussian with a second-order polynomial. For the BIS measurements, we used the bisector fits provided by the same pipeline, although this time series did not provide any relevant information for the stellar activity study since the BIS is not well defined for M-dwarfs due to the bumps present on the wings of the CCF \citep{2020ExA....49...73R}.

\begin{figure}
        \includegraphics[width=\columnwidth]{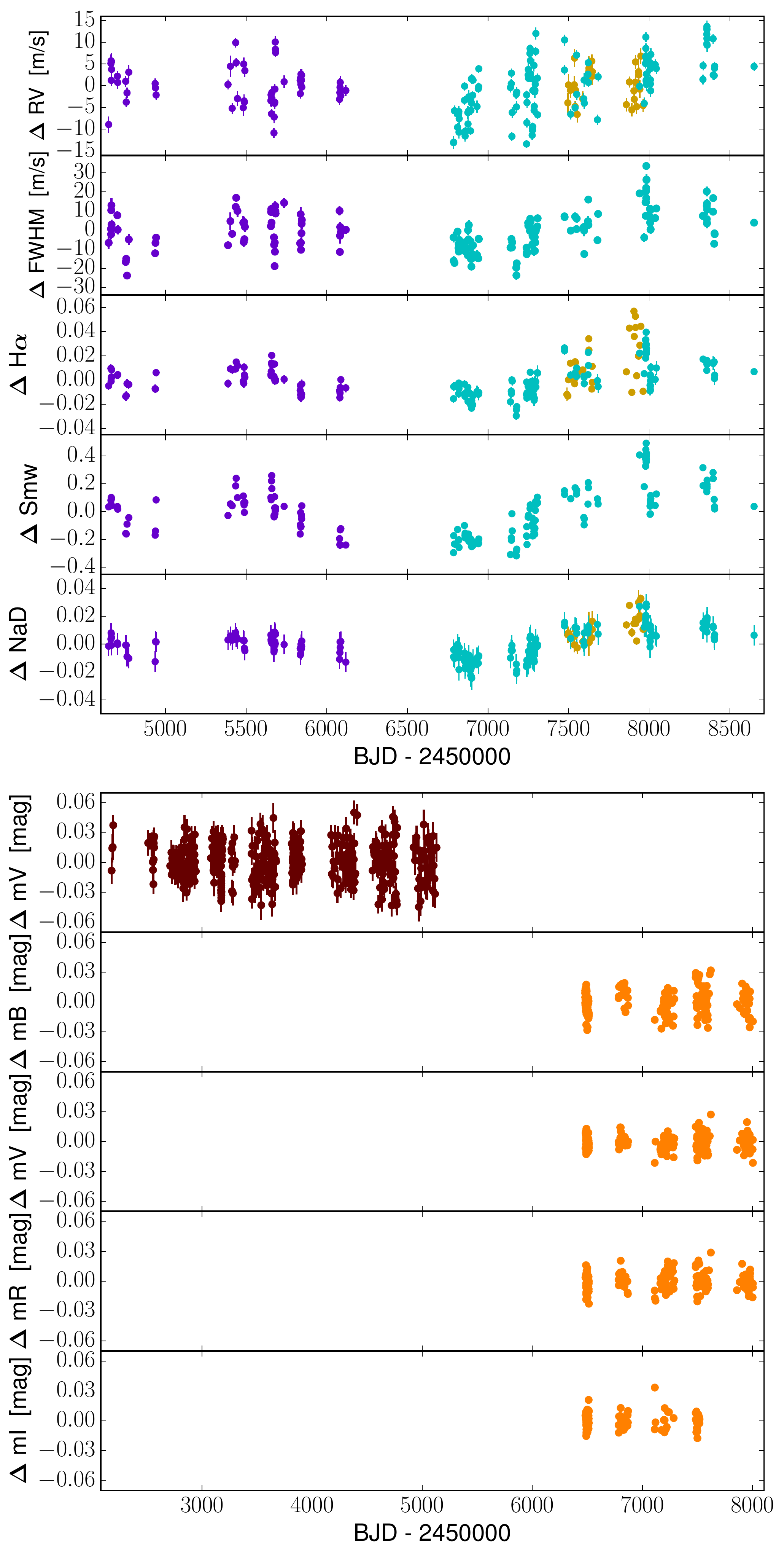} 
    \caption{\textbf{Top:} Time series of the RV and the chromospheric activity indicators with combined measurements from HARPS (represented in violet), HARPS-N (represented in cyan), and CARMENES (represented in yellow). \textbf{Bottom:} Time series of the photometric magnitudes taken with ASAS (represented in dark red) and EXORAP (represented in orange) in the $B$, $V$, $R$, and $I$ filters.}
    \label{fig:Compact_Indexes}
\end{figure}

To remove outliers in all time series, we applied a 3$\sigma$-clipping in values and another 3$\sigma$-clipping in errors to each one of them, using the median value and error as a reference. This serves to discard points that can be related to stellar flares, a phenomenon that is especially important in the case of active M-type stars \citep{2009A&A...498..853R,2014ApJ...797..121H}. For consistency, a measurement cataloged as an outlier in one of the time series was discarded in the rest of the time series. The final measurements resulting from this process are shown in Fig.~\ref{fig:Compact_Indexes}, along with the photometric time series, and the properties of each dataset are listed in Table~\ref{tab:Indexes_Properties}.

\begin{table}
\centering
        \caption{Properties of all the datasets used in this work.}
        \label{tab:Indexes_Properties}
        \begin{tabular}{lcccc}
                \hline
                Index & Mean Value & RMS & Mean Error \\
                \hline
                RV [m\,s$^{-1}$]     & 10619  & 6.166  & 1.209 \\
                FWHM [m\,s$^{-1}$]   & 4505   & 10.35  & 1.978 \\
                H${\rm \alpha}$      & 0.386  & 0.015  & 0.003 \\
                S$_{\rm mw}$         & 1.599  & 0.181  & 0.009 \\
                NaD                  & 0.426  & 0.011  & 0.007 \\
                m$_{\rm V}$ (ASAS)   & 9.224  & 0.019  & 0.012 \\
                m$_{\rm B}$ (EXORAP) & 4.723  & 0.015  & 0.0011 \\
                m$_{\rm V}$ (EXORAP) & 3.448  & 0.010  & 0.0016 \\
                m$_{\rm R}$ (EXORAP) & 2.279  & 0.0094 & 0.0013 \\
                m$_{\rm I}$ (EXORAP) & 1.550  & 0.011  & 0.0011 \\
                \hline
        \end{tabular}
        \begin{minipage}{\columnwidth}
\end{minipage}  
\end{table}

\section{Time-series analysis}

\label{sec:Analysis}

\begin{figure}
        \includegraphics[width=\columnwidth]{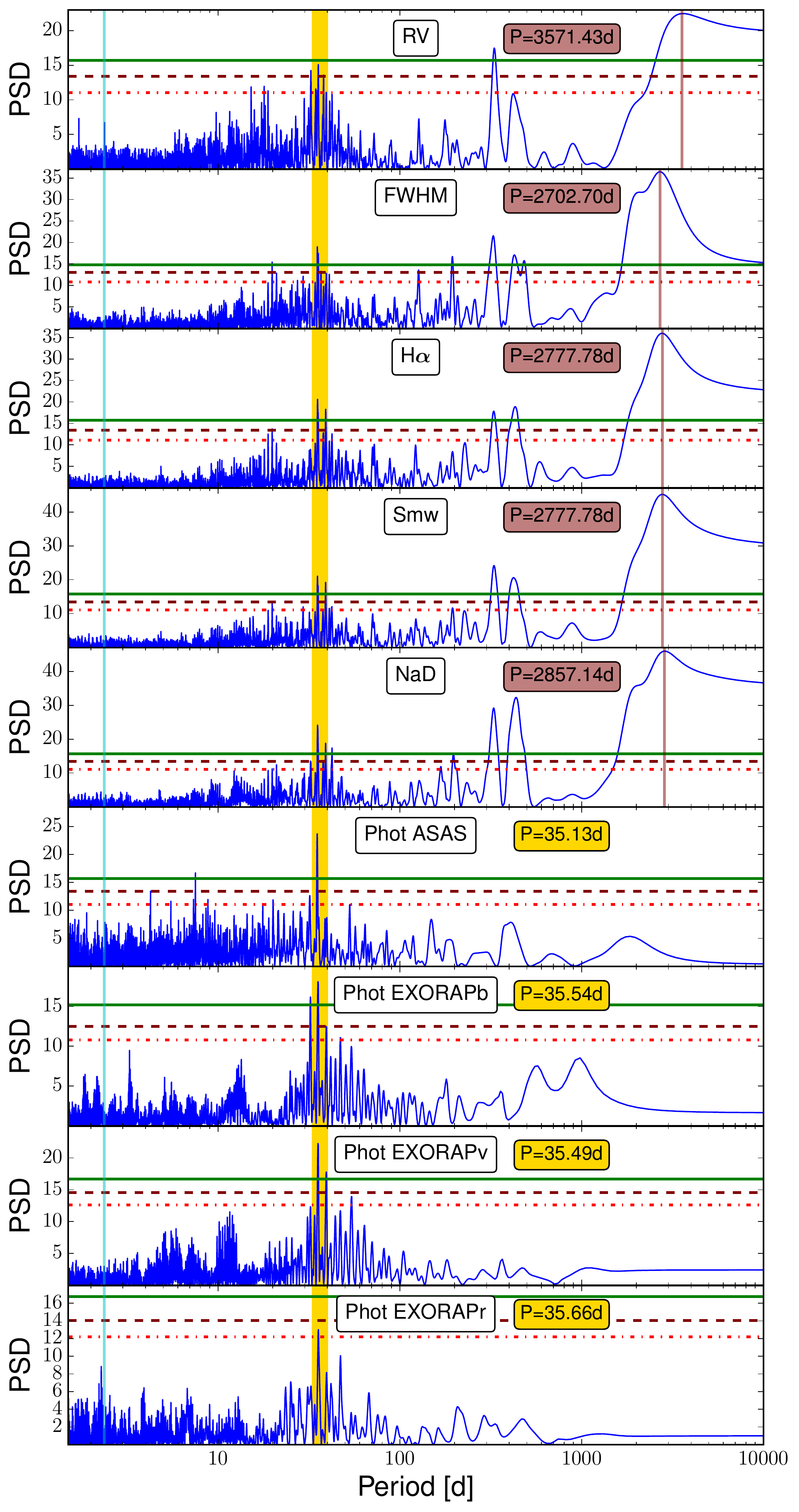} 
    \caption{Periodograms of the RV, FWHM, H${\rm \alpha}$, S$_{\rm mw}$, NaD, ASAS $V$, EXORAP $B$, EXORAP $V$, and EXORAP $R$-band magnitude time series. The periods related to stellar rotation, planetary companion, and possible magnetic cycle have been represented as yellow, cyan, and brown vertical lines,  respectively. The green, brown, and red horizontal lines indicate the 0.1\%, 1\%, and 10\% FAP levels, respectively.}
    \label{fig:Compact_FirstPeriodograms}
\end{figure}

We analyzed the periodic signals present in the time series of RVs, photometric observations, and chromospheric activity indicators related to stellar rotation and magnetic activity. First, we built the generalized Lomb Scargle (GLS) periodograms \citep{1976Ap&SS..39..447L,2009A&A...496..577Z} and established three threshold levels related to the false alarm probability (FAP) of these signals \citep{1986ApJ...302..757H}. The threshold levels were obtained by randomizing each time series separately in a bootstrapping process of 10000 iterations \citep{2001A&A...374..675E}. We then studied the most significant peaks in the periodograms based on these threshold levels. For this study, we adopted the significance standards established in \citet{2019MNRAS.488.5145T}, with the 0.1\% and 10\% levels FAP separating the statistically significant signals, those whose significance is unclear, and the non-significant ones. The first periodograms from the stellar activity indicators, along with the photometric data and the RVs are shown in Fig.~\ref{fig:Compact_FirstPeriodograms}.

Using the S-index values, we computed a mean chromospheric activity level of $\log_{10} ($R$_{\rm HK}^{'})$~=~$-$4.88 $\pm$ 0.05, which matches the one published in \citet{2018A&A...612A..89S}. That publication presents an analysis of the whole HADES sample to establish a relation between the activity level and the rotation period of the stars of the sample, including GJ~740, which we found to have a $P_{\rm rot}$~=~36.3~$\pm$~1.7\,d. Figure~\ref{fig:Compact_FirstPeriodograms} clearly shows that all the periodograms present significant peaks near to this value (yellow line), along with two bumps at 330\,d and 440\,d. In all the spectroscopic periodograms, the statistical significance of the rotation signal is exceeded by a long-period signal that could be associated with the magnetic cycle of the star.

\subsection{Long-term variation}

\begin{figure} 
        \includegraphics[width=\columnwidth]{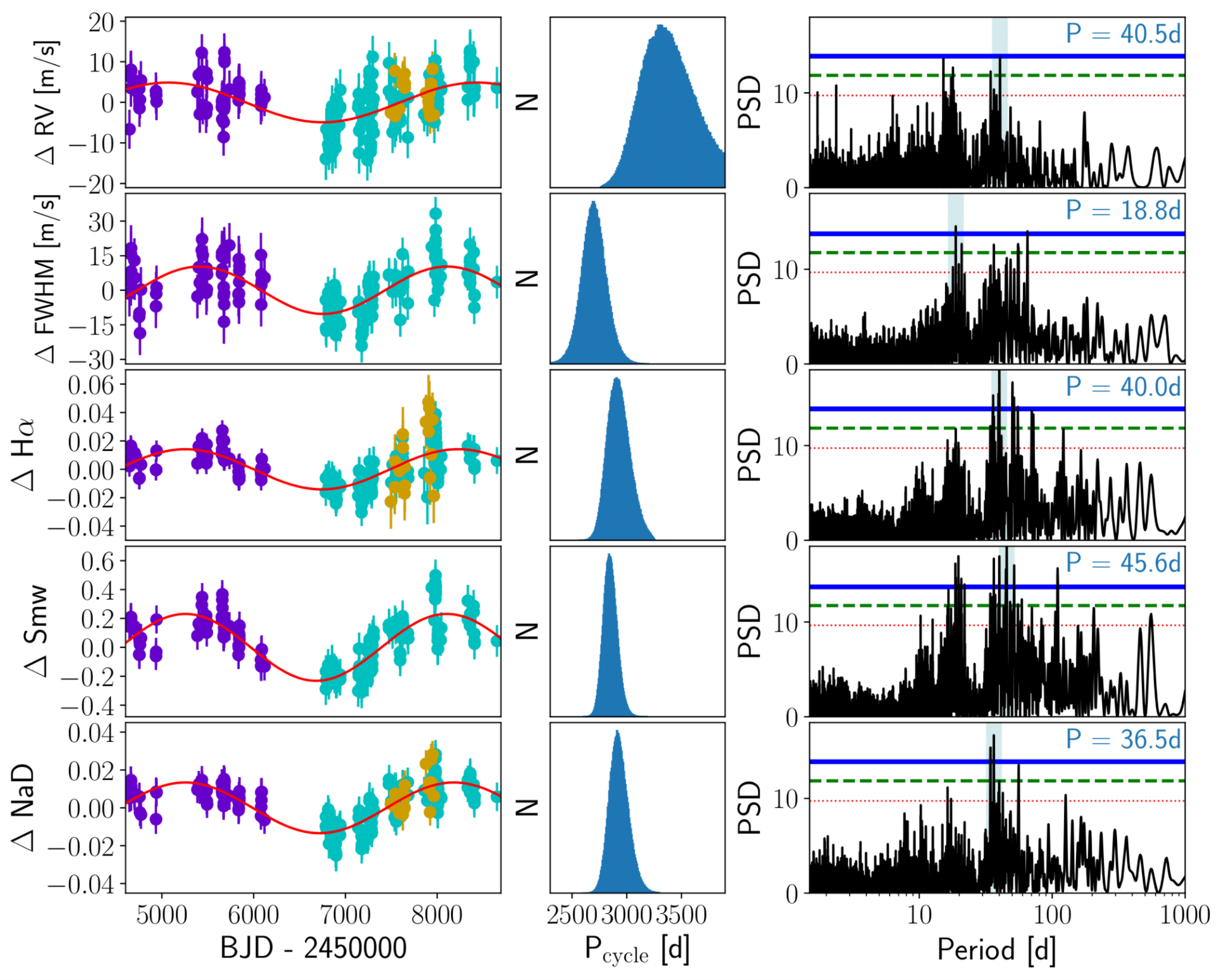} 
    \caption{Analysis of the long-term signal. \textbf{Left:} RV, FWHM, H${\rm \alpha}$, S$_{\rm mw}$, and NaD time series with their respective model of the long-term signal. \textbf{Center:} Posterior distribution for the periodicity of the long-term signal. \textbf{Right:} Periodograms of the residuals after subtracting the model. The blue, green, and red horizontal lines indicate the 0.1\%, 1\%, and 10\% FAP levels, respectively. The highest peak of the periodogram is marked with a blue shaded area.}
    \label{fig:6_2_Cycle}
\end{figure}

The most significant peak in all the spectroscopic GLS periodograms of Fig.~\ref{fig:Compact_FirstPeriodograms} is related to a long-term signal whose periodicity around 8\,yr makes it compatible with a magnetic cycle similar to the one observed in the Sun. The length of the photometric campaigns is not enough to detect a signal with this periodicity. This signal has a significance much greater than the 0.1\%~FAP level in all cases. We fitted this signal with a sinusoidal model that includes offset and jitter terms for each spectrograph. We first used an implementation of the Broyden–Fletcher–Goldfarb–Shanno (BFGS) algorithm \citep{schraudolph2007stochastic} available in the \texttt{minimize} package from the \texttt{scipy} library to optimize the parameters of the model. Then we carried out an Markov Chain Monte Carlo (MCMC) analysis based on the results of the BFGS algorithm using the \texttt{emcee} package \citep{2013PASP..125..306F}, performing simulations within a Bayesian framework to infer the probability distribution over all the parameters considered. We used 10\,000 steps for the burn-in stage, 50\,000 steps for the construction stage, and 512 walkers to sample the parameter space. We established a convergence criterion based on the auto-correlation function to ensure that the previous setup produces a correct parameter distribution. This analysis provided the results shown in Fig.~\ref{fig:6_2_Cycle} when the long-term component is modeled along with the rotation component.

In the middle panels of Fig.~\ref{fig:6_2_Cycle}, we show how the long-term signal in the RV time series presents a shift in periodicity with respect to the signals detected in the activity indicators, displaying a compatibility with them at 2$\sigma$. The periodicity of the signal in the activity indices is around 2820\,d (2701$^{+110}_{-107}$ in FWHM, 2832$^{+105}_{-94}$ in H${\rm \alpha}$, 2848$^{+65}_{-59}$ in S$_{\rm mw}$, and 2913$^{+91}_{-81}$ in NaD), while in the RV time-series is located at 3363$^{+252}_{-215}$\,d. Our time span of~$\sim$~4000\,d did not allow us to detect a second cycle to ensure the periodicity of the signal.

The right panels of Fig.~\ref{fig:6_2_Cycle} show the periodograms related to the residuals obtained after subtracting the models represented in the left panels from each time series. The jitter terms provided by the MCMC were added quadratically to the errors of each dataset and the offset applied to all the measurements. In these plots, the bumps at 330\,d and 440\,d from Fig.~\ref{fig:Compact_FirstPeriodograms} have completely disappeared from the periodograms, which indicates that these signals were aliases of the long-term signal. We found signals related to the rotation of the star around 40\,d in the RV, H${\rm \alpha}$, S$_{\rm mw}$, and NaD time series as the main peaks. In the FWHM case, the signal related to the first harmonic of the stellar rotation shows a greater amplitude than the forest of peaks around 40\,d.

\subsection{Rotation period}

\begin{figure}
        \includegraphics[width=\columnwidth]{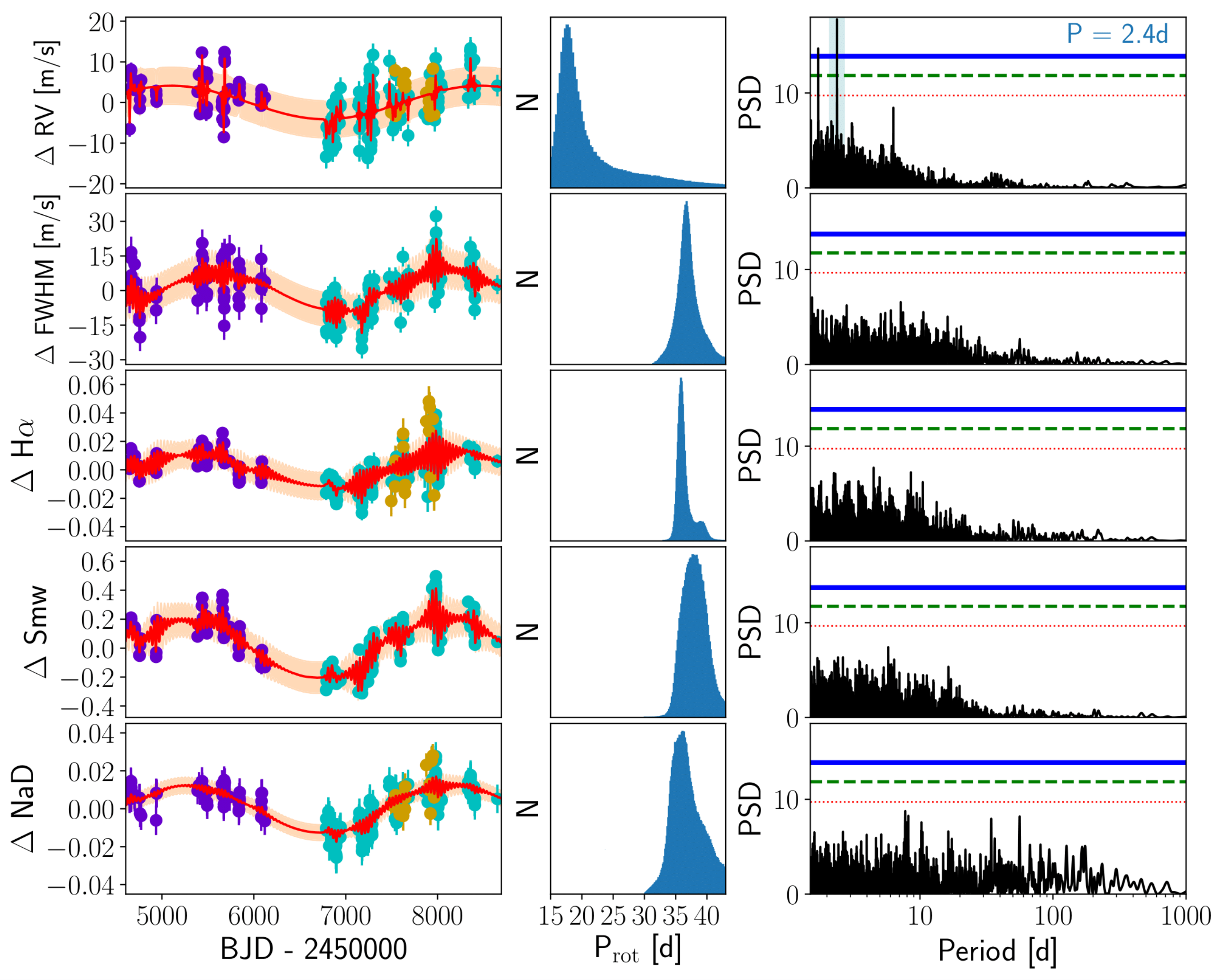}
    \caption{Analysis of the rotation signal. \textbf{Left:} RV, FWHM, H${\rm \alpha}$, S$_{\rm mw}$, and NaD time series with their respective model of the long-term signal and the rotation signal. The shaded regions indicate the 1$\sigma$ confidence band of the GP model. \textbf{Center:}  Posterior distribution for the periodicity of the rotation signal. \textbf{Right:} Periodograms of the residuals after subtracting the model. The blue, green, and red horizontal lines indicate the 0.1\%, 1\%, and 10\% FAP levels, respectively. The highest peak of the periodogram is marked with a blue shaded area.}
    \label{fig:6_3_Rotation}
\end{figure}

For the next step of the analysis, we used the original time series to characterize the rotation through a Gaussian process (GP) regression \citep{2014MNRAS.443.2517H,2015MNRAS.452.2269R,2015ITPAM..38..252A} and to simultaneously recompute the activity cycle model that was used previously. To model the stellar rotation, we used the following quasi-periodic kernel:

\begin{equation}
  \kappa(\tau)=\frac{K_{\rm rot}^{2}}{2+C}e^{-\tau/t_{s}}\left[\cos\left(\frac{2 \pi \tau}{P_{\rm rot}}\right)+(1+C)\right]+\left(\sigma_{\rm RV}^{2}(t)+\sigma_{\rm j}^{2}\right)\delta_{\tau}
  \label{eq:Kernel}
,\end{equation}

\noindent which contains the squared amplitude of the signal $K_{\rm rot}^{2}$, the periodicity of the signal $P_{\rm rot}$, and the timescale of the surface phenomena in the star $t_{s}$. The kernel also contains a term named $C$ whose role is to measure the relative importance between the two components of the kernel: the periodic (i.e., the cosine) and non-periodic (i.e., the exponential). This kernel has been proven to provide good results in the activity analysis of Proxima Centauri performed by \citet{2020A&A...639A..77S}. We fitted the parameters of the kernel using the \texttt{celerite} code \citep{2017AJ....154..220F} with the same setup of steps, walkers, and convergence criterion as the one described in the previous subsection. The parameters obtained for all of the time series (including the new values of the parameters related to the long-term signal) are listed in Table~\ref{tab:CycleRotPlanet_Properties} and the results are shown in Fig.~\ref{fig:6_3_Rotation}.

\renewcommand{\arraystretch}{1.5}

\begin{table*}
\centering
        \caption{Priors and parameters related to the long-term and rotation signals obtained from the final stellar activity MCMC analysis of all the time series.}
        \label{tab:CycleRotPlanet_Properties}
        \begin{tabular}{lccccc}
                \hline
                Parameter & RV & FWHM & H${\rm \alpha}$ & S$_{\rm mw}$ & NaD \\
                \hline
                \multicolumn{6}{c}{Cycle priors} \\
                \hline
                $K_{\rm cycle}$     & $\mathcal{U}$ (0.01,     20.0) & $\mathcal{U}$ (0.01,     20.0) & $\mathcal{U}$ (10$^{-6}$, 0.02)
                                    & $\mathcal{U}$ (10$^{-6}$, 0.4) & $\mathcal{U}$ (10$^{-6}$, 0.4) \\
                $P_{\rm cycle}$ [d] & $\mathcal{U}$ (2300.0, 3900.0) & $\mathcal{U}$ (2300.0, 3900.0) & $\mathcal{U}$ (2300.0,  3900.0) 
                                    & $\mathcal{U}$ (2300.0, 3900.0) & $\mathcal{U}$ (2300.0, 3900.0) \\
                $T$ [d]             & $\mathcal{U}$ (2500.0, 3600.0) & $\mathcal{U}$ (2500.0, 3600.0) & $\mathcal{U}$ (2500.0,  3600.0)
                                    & $\mathcal{U}$ (2500.0, 3600.0) & $\mathcal{U}$ (2500.0, 3600.0) \\
                \hline
                \multicolumn{6}{c}{Cycle values} \\
                \hline
                $K_{\rm cycle}$     & 4.2$^{+1.1}_{-1.0}$\,m\,s$^{-1}$ & 8.3$^{+2.3}_{-2.3}$\,m\,s$^{-1}$ & 0.0113$^{+0.0036}_{-0.0036}$ 
                                    & 0.205$^{+0.037}_{-0.039}$        & 0.1247$^{+0.0018}_{-0.0019}$     \\
                $P_{\rm cycle}$ [d] & 3363$^{+230}_{-217}$             & 2640$^{+268}_{-296}$             & 2860$^{+203}_{-204}$
                                    & 2849$^{+178}_{-146}$             & 2877$^{+206}_{-186}$             \\
                $T$ [d]             & 2976$^{+205}_{-182}$             & 2860$^{+134}_{-118}$             & 2818$^{+114}_{-127}$
                                    & 2751$^{+95}_{-80}$               & 2801$^{+92}_{-76}$               \\
                \hline
                \multicolumn{6}{c}{Rotation priors} \\
                \hline
                $K_{\rm rot}^{2}$ & $\mathcal{LU}$ (8.0,              40.0) & $\mathcal{LU}$ (20.0,    100.0) & $\mathcal{LU}$ (10$^{-6}$, 0.1)  
                                  & $\mathcal{LU}$ (4$\times10^{-5}$, 0.03) & $\mathcal{LU}$ (10$^{-6}$, 0.1) \\
                $P_{\rm rot}$ [d] & $\mathcal{LU}$ (15.0,  45.0)            & $\mathcal{LU}$ (15.0, 45.0)     & $\mathcal{LU}$ (15.0, 45.0)      
                                  & $\mathcal{LU}$ (15.0,  45.0)            & $\mathcal{LU}$ (15.0, 45.0)     \\
                $t_{\rm s}$ [d]   & $\mathcal{LU}$ (1.0, 300.0)             & $\mathcal{LU}$ (1.0, 300.0)     & $\mathcal{LU}$ (1.0, 300.0) 
                                  & $\mathcal{LU}$ (1.0, 300.0)             & $\mathcal{LU}$ (1.0, 300.0)     \\
                $C$      & $\mathcal{LU}$ (0.0, 1.0)               & $\mathcal{LU}$ (0.0, 1.0)       & $\mathcal{LU}$ (0.0,   1.0)
                                  & $\mathcal{LU}$ (0.0, 1.0)               & $\mathcal{LU}$ (0.0, 1.0)       \\
                \hline
                \multicolumn{6}{c}{Rotation values} \\
                \hline
                $K_{\rm rot}$ $^{(*)}$ & 4.5$^{+2.0}_{-1.8}$\,m\,s$^{-1}$ & 7.4$^{+4.3}_{-3.7}$\,m\,s$^{-1}$   & 0.0102$^{+0.0057}_{-0.0050}$                        
                                       & 0.108$^{+0.059}_{-0.050}$        & 0.0043$^{+0.0030}_{-0.0026}$       \\
                $P_{\rm rot}$ [d]      & 18.6$^{+6.0}_{-1.8}$             & 36.9$^{+2.0}_{-1.4}$               & 36.1$^{+1.9}_{-0.7}$
                                       & 37.1$^{+2.0}_{-1.9}$             & 36.5$^{+3.0}_{-2.0}$               \\
                $t_{\rm s}$ [d]        & 12.0$^{+3.9}_{-1.5}$             & 93$^{+82}_{-49}$                   & 170$^{+76}_{-66}$
                                       & 92$^{+50}_{-36}$                 & 109$^{+58}_{-51}$                  \\
                $\log C$               & $-$21$^{+14}_{-14}$              & $-$21$^{+14}_{-13}$                & $-$21$^{+14}_{-14}$
                                       & $-$21$^{+14}_{-14}$              & $-$20$^{+14}_{-14}$                \\
                \hline
                \multicolumn{6}{c}{Remaining priors} \\
                \hline
                jitter$_{\rm HARPS}$    & $\mathcal{LU}$ (0.01,   4.0)  & $\mathcal{LU}$ (3.0, 9.0)         & $\mathcal{LU}$ (0.001, 0.02) 
                                        & $\mathcal{LU}$ (0.01,  0.10)  & $\mathcal{LU}$ (10$^{-4}$, 0.008) \\
                jitter$_{\rm HARPS-N}$  & $\mathcal{LU}$ (0.01,   4.0)  & $\mathcal{LU}$ (3.0, 9.0)         & $\mathcal{LU}$ (0.001, 0.02)      
                                        & $\mathcal{LU}$ (0.01,  0.10)  & $\mathcal{LU}$ (10$^{-4}$, 0.008) \\
                jitter$_{\rm CARMENES}$ & $\mathcal{LU}$ (0.01,   4.0)  & $\cdots$                          & $\mathcal{LU}$ (0.001, 0.02)      
                                        & $\cdots$                      & $\mathcal{LU}$ (10$^{-4}$, 0.008) \\
                offset$_{\rm HARPS}$    & $\mathcal{U}$ ($-$15.0, 15.0) & $\mathcal{U}$ ($-$80.0, 80.0)     & $\mathcal{U}$ ($-$0.2,   0.2) 
                                        & $\mathcal{U}$ ($-$0.4,   0.4) & $\mathcal{U}$ ($-$0.2,   0.2)     \\
                offset$_{\rm HARPS-N}$  & $\mathcal{U}$ ($-$15.0, 15.0) & $\mathcal{U}$ ($-$80.0, 80.0)     & $\mathcal{U}$ ($-$0.2,   0.2)
                                        & $\mathcal{U}$ ($-$0.4,   0.4) & $\mathcal{U}$ ($-$0.2,   0.2)     \\
                offset$_{\rm CARMENES}$ & $\mathcal{U}$ ($-$15.0, 15.0) & $\cdots$                          & $\mathcal{U}$ ($-$0.2,   0.2)
                                        & $\cdots$                      & $\mathcal{U}$ ($-$0.2,   0.2)     \\
                \hline
                \multicolumn{6}{c}{Remaining values} \\
                \hline
                jitter$_{\rm HARPS}$    & 0.20$^{+0.85}_{-0.17}$\,m\,s$^{-1}$ & 5.72$^{+0.93}_{-0.80}$\,m\,s$^{-1}$ & 0.00153$^{+0.00013}_{-0.00012}$
                                        & 0.0403$^{+0.0078}_{-0.0067}$        & 0.00144$^{+0.00066}_{-0.00033}$     \\
                jitter$_{\rm HARPS-N}$  & 2.47$^{+0.36}_{-0.34}$\,m\,s$^{-1}$ & 3.76$^{+0.45}_{-0.39}$\,m\,s$^{-1}$ & 0.00380$^{+0.00052}_{-0.00047}$
                                        & 0.0342$^{+0.0039}_{-0.0036}$        & 0.00129$^{+0.00045}_{-0.00022}$     \\
                jitter$_{\rm CARMENES}$ & 0.13$^{+0.69}_{-0.11}$\,m\,s$^{-1}$ & $\cdots$                            & 0.01008$^{+0.00027}_{-0.00020}$
                                        & $\cdots$                            & 0.0049$^{+0.0015}_{-0.0015}$        \\
                offset$_{\rm HARPS}$    & $-$5.5$^{+1.3}_{-1.3}$\,m\,s$^{-1}$ & 6.6$^{+3.1}_{-3.1}$\,m\,s$^{-1}$    & $-$0.0098$^{+0.0042}_{-0.0041}$
                                        & $-$0.128$^{+0.039}_{-0.038}$        & $-$0.0041$^{+0.0020}_{-0.0020}$     \\
                offset$_{\rm HARPS-N}$  & 2.3$^{+1.2}_{-1.1}$\,m\,s$^{-1}$    & $-$4.9$^{+2.2}_{-2.1}$\,m\,s$^{-1}$ & 0.0066$^{+0.0034}_{-0.0034}$
                                        & 0.084$^{+0.033}_{-0.032}$           & $-$0.0031$^{+0.0018}_{-0.0017}$     \\
                offset$_{\rm CARMENES}$ & $-$1.0$^{+1.9}_{-1.9}$\,m\,s$^{-1}$ & $\cdots$                            & $-$0.0082$^{+0.0060}_{-0.0063}$
                                        & $\cdots$                            & $-$0.0073$^{+0.0033}_{-0.0032}$     \\
                \hline
        \end{tabular}
        \begin{minipage}{17.5 cm}
{\footnotesize $^{(*)}$ The K$_{\rm rot}$ values were calculated as the root square of the $K_{\rm rot}^{2}$ posterior distribution results.}
\end{minipage}  
\end{table*}

\begin{figure}
        \includegraphics[width=\columnwidth]{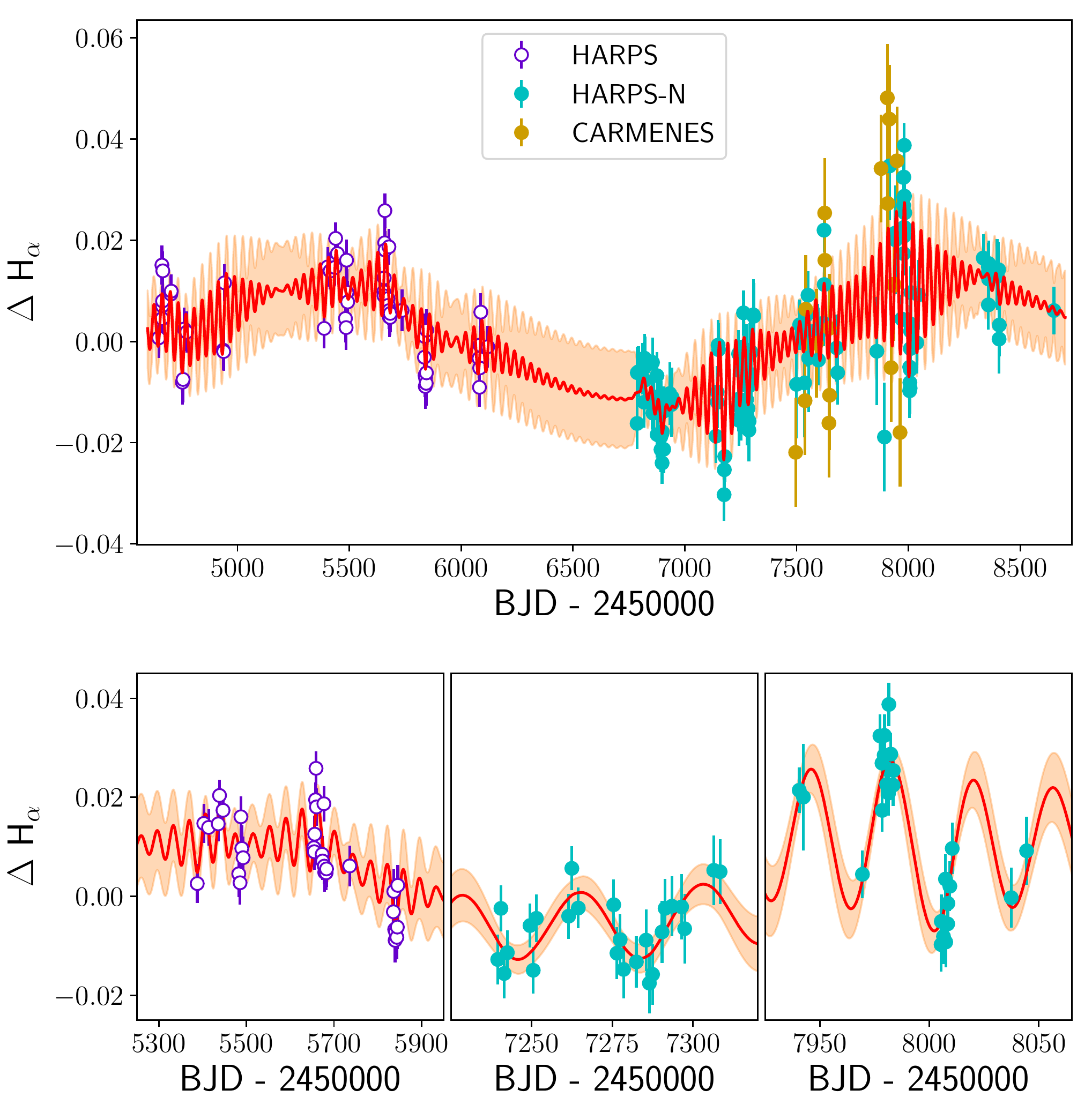}
    \caption{\textbf{Top:} Model obtained for the H${\rm \alpha}$ time series using Gaussian Processes to treat the rotation signal and a sinusoidal function to treat the long-term signal. The shaded regions indicate the 1$\sigma$ confidence band of the GP model. \textbf{Bottom:} Zoom on different time windows.}
    \label{fig:6_4_Rotation_Model}
\end{figure}

The middle panels of Fig.~\ref{fig:6_3_Rotation} show that the rotation period in the RV time series is not well-constrained, with a wide distribution reaching its maximum peak close to the first harmonic of the rotation period. The coherence time recovered from the RV kernel is much shorter than the one obtained in the rest of the time series. In the chromospheric activity-indicator time series, we detected the rotation signal with a periodicity of~$\sim$~37\,d (36.9$^{+2.0}_{-1.4}$ in FWHM, 36.1$^{+1.9}_{-0.7}$ in H${\rm \alpha}$, 37.1$^{+2.0}_{-1.9}$ in S$_{\rm mw}$, and 36.5$^{+3.0}_{-2.0}$ in NaD). The fact that the parameter $t_{\rm s}$ is a few stellar rotations in the activity indices is consistent with the evolutionary timescale of the active regions \citep{2017A&A...598A..28S}. Of these activity indicators, the H${\rm \alpha}$ index exhibits the most significant stellar rotation signal (see Fig.~\ref{fig:6_2_Cycle}) and provides the most stable stellar activity characterization, with lower relatives errors in the rotation parameters fitted (see Table~\ref{tab:CycleRotPlanet_Properties}). A zoom-in of the H${\rm \alpha}$ model is shown in Fig.~\ref{fig:6_4_Rotation_Model}.

We then computed the Bayesian evidence $\log Z$ \citep{perrakis2014use} for all the time series. When comparing the $\log Z$ values of two different models, a difference between their $\log Z$ values greater than~10 indicates a significant preference for the model with the higher $\log Z$. The results indicate that the rotation+long-term signal model is preferred over the long-term signal model. The first model is characterized by greater $\log Z$ values in all the time series, presenting a difference greater than 30 with respect to the values computed for the second model. This Bayesian parameter is shown in Table~\ref{tab:Bayessian_Properties} for all the models considered in this work.

\begin{table*}
\centering
        \caption{$\rm \log Z$ values computed for the different models implemented in this paper within their corresponding time series.}
        \label{tab:Bayessian_Properties}
        \begin{tabular}{lcccccccccc}
                \hline
                Model                           & RV       & FWHM     & H${\rm \alpha}$ & S$_{\rm mw}$ & \multicolumn{2}{c}{NaD}              \\
                \hline
                Cycle                           & $-$590   & $-$632   & 625             & 141          & 672      \\
                Cycle+Rotation                  & $-$555   & $-$601   & 683             & 208          & 721      \\
                Cycle+Rotation+Keplerian Planet & $-$533   & $\cdots$ & $\cdots$        & $\cdots$     & $\cdots$ \\
                Cycle+Rotation+Circular Planet  & $-$535   & $\cdots$ & $\cdots$        & $\cdots$     & $\cdots$ \\
                \hline
        \end{tabular}
        \begin{minipage}{17.5cm}
\end{minipage}  
\end{table*}

In Fig.~\ref{fig:Compact_FirstPeriodograms} we previously observed how the ASAS photometric dataset exhibit the same rotation signal as the one found in the chromospheric time series. The GP regression with the rotation kernel reveals a periodicity of $P_{\rm rot}$=35.60$^{+0.89}_{-0.55}$\,d for this signal. No additional significant signals were detected after this model was subtracted.

The analysis of the SuperWASP photometric time series did not reveal any clear information about the rotation of GJ~740 since all the short-term signals have low significance. This time series presents the problem that results from the majority of its photometric magnitudes having been  measured in a time span of 130\,d, with only 9\,d of observations outside this range. The time span of the dataset and the density of points is not enough to have a good characterization of the long-term behavior of the star, a fact that is reflected in the lack of long-term significant peaks after the trend correction. We found a similar time span problem in the APACHE dataset analyzed in \citet{2020MNRAS.491.5216G} with regard to the long-term signal. 

We also carried out a pre-whitening analysis of the four EXORAP light curves. The main feature of the periodograms of the pre-whitened $B$ and $V$ light curves is a forest of strong peaks between 20 and 60 days, which are aliases of the strongest peak at~$\sim$~36 days with FAP~$<$~0.1\%. The same applies to the $R$-band light curve, where the 36 d period has a slightly higher FAP~$\sim$~1\%. In the $I$-band all the signals are non-statistical significant (i.e., FAP~$>$~10\%) due to the short time span of the light curve.

We then performed a more sophisticated analysis by fitting the light curves with a cubic polynomial function in a Monte Carlo framework, using the \texttt{emcee} package \citep{2013PASP..125..306F}. We included a GP to characterize the red noise in the data introduced by stellar activity, allowing independent offsets for the two different instrumental setups used during the survey. The posterior distribution of the $B$, $V$, and $R$ light curves shows that the correlated noise is consistent with the periodicity of 36 days returned by the periodograms. The combination of these distributions provides an orbital period of 35.563~$\pm$0.071\,d. The fit of the $I$-band light curve does not lead to any conclusive result. This is due to the fact that this light curve is shorter in time coverage, contains fewer data, and the activity signal in this red band is expected to be lower than in the previous cases.

The periodograms of the stellar indices residuals in the right panels of Fig.~\ref{fig:6_3_Rotation} show that following the rotation subtraction, no more signals were detected with a statistical significance higher than the 10\% FAP level (except in the RV case). This indicates that the signals previously detected at~$\sim$~19\,d were related to the first harmonic of the rotation period. In the case of the RV time series, we found a short-period signal of 2.4\,d with a significance greater than the 0.1\% level of FAP, which could have a planetary origin since it is not present in the activity proxy time series.

\subsection{Planetary signal}

\begin{figure*}
    \centering
        \includegraphics[width=14.5cm,page=1]{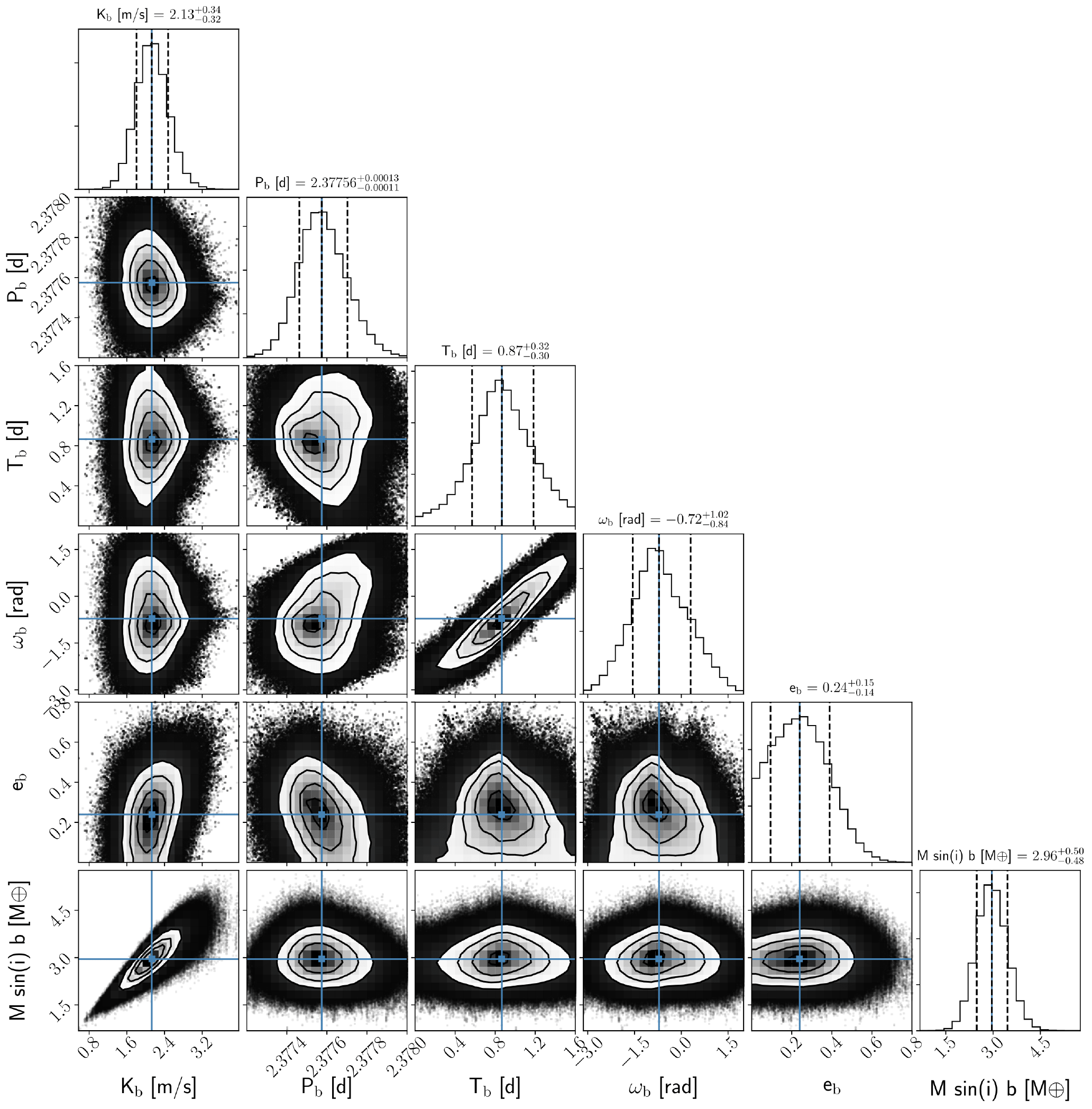}
    \caption{Posterior distributions of the GJ~740~b fitted parameters. The 16th-84th percentiles are represented through vertical dashed lines.}
    \label{fig:RV_MCMC_SineRot1Pkepler_fixRot_1}
\end{figure*}

We explored the possible presence of a planetary signal at 2.38\,d in the RV time series by adding a Keplerian component to our previous MCMC model that included the stellar rotation and long-term signal terms. We performed an MCMC analysis using the H${\rm \alpha}$ model shown in Fig.~\ref{fig:6_4_Rotation_Model} to establish the boundaries for the RV rotation parameters. The Keplerian parameters of the candidate planet converged to the distributions shown in Fig.~\ref{fig:RV_MCMC_SineRot1Pkepler_fixRot_1}. The rest of the parameters are displayed in Table~\ref{tab:CycleRot_Properties} and Fig.~\ref{fig:RV_MCMC_SineRot1Pkepler_fixRot_3}.

\begin{figure}
        \includegraphics[width=\columnwidth,page=1]{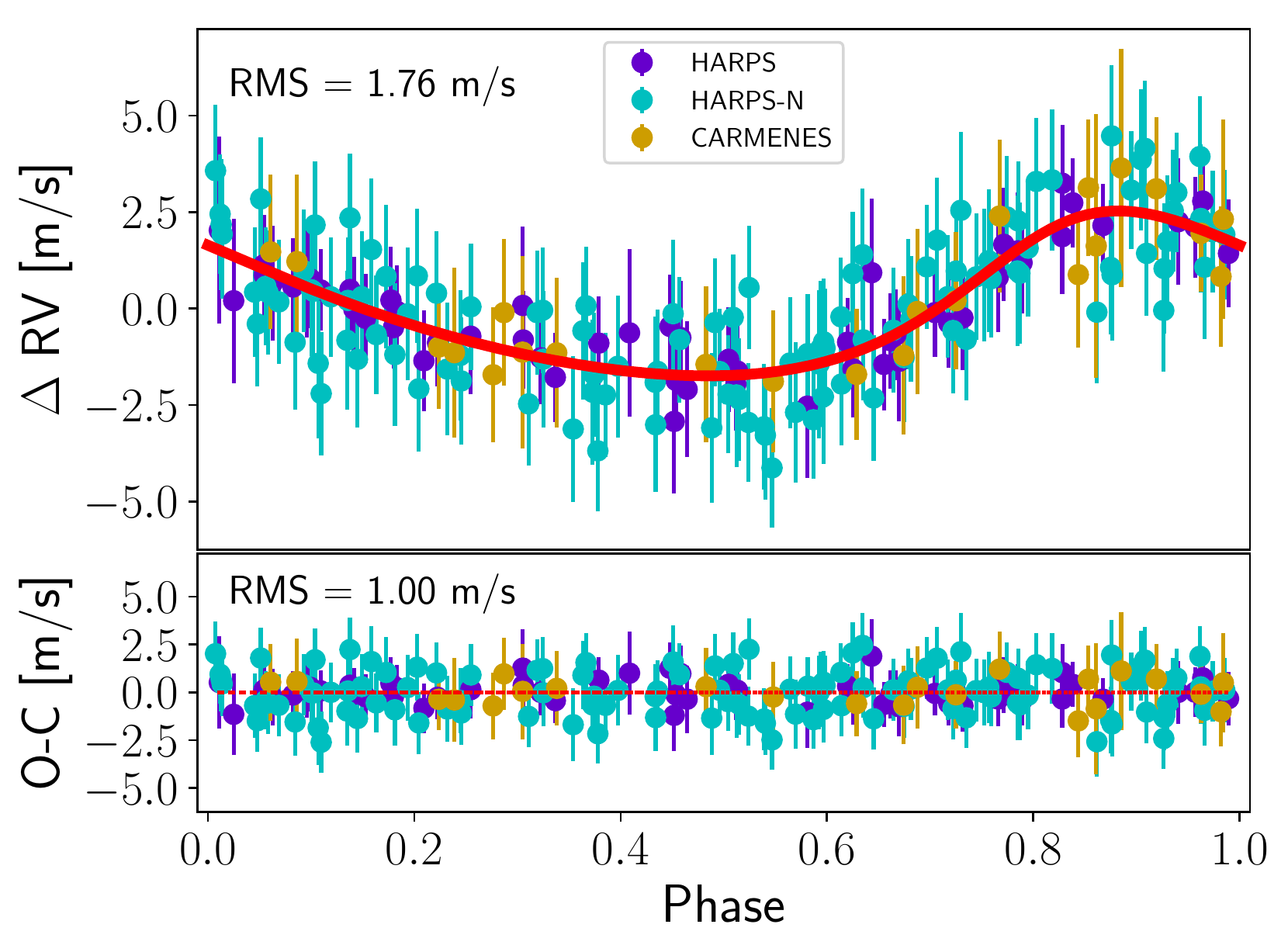}
    \caption{\textbf{Top:} Phase-folded curve of the RV time series using the GJ~740~b orbital period after subtracting the long-term signal and the rotation period. The jitter terms of each spectrograph have been added to the original RV errors. The red solid line represents the planetary model. \textbf{Bottom:} Residuals after subtracting the model.}
    \label{fig:RV_MCMC_SineRot1Pkepler_fixRot_2}
\end{figure}

The posterior distributions from this MCMC analysis exhibit a good convergence based on the auto-correlation of the chains to an orbital period of 2.37756$^{+0.00013}_{-0.00011}$\,d for GJ~740~b. The signal is characterized by an amplitude of 2.13$^{+0.34}_{-0.32}$\,m\,s$^{-1}$ and an eccentricity of 0.24$^{+0.15}_{-0.14}$. The $\log Z$ value computed for this model indicates a significantly better Bayesian result than the one obtained for the previous models implemented for the RV time series, with a $\Delta\log Z$ greater than~20 in favor of the planetary model. The planetary nature of this signal is further supported by the steady increase of its statistical significance and the consistency of the RV semi-amplitude with the number of measurements shown in Fig.~\ref{fig:FAP_Evolution}. Fig.~\ref{fig:RV_MCMC_SineRot1Pkepler_fixRot_2} depicts the RV time series phased to the period of this planetary signal.

The nominal eccentricity obtained is larger than the one expected for a short orbital period planet such as GJ~740~b, but it is consistent with zero at the 2$\sigma$ level. For this reason, we tried an additional model using a sinusoidal function to represent the planetary signal. This model is characterized by a Bayesian evidence of $\log Z$=$-$535, which is slightly worse than the value obtained for the Keplerian model. The difference between the two models is below the limit to consider one of them more significant than the other.

Using the orbital period obtained we computed the semi-axis of the planet using the mass of the host star from Table~\ref{tab:GJ740_Properties}. This parameter allowed us to calculate the flux received by the planet, its equilibrium temperature, and the probability that the planet could transit its host star. We listed in Table~\ref{tab:GJ740b_Properties} all the MCMC parameters, along with these derived properties.

\begin{table}
\centering
        \caption{MCMC and derived planetary parameters of GJ~740~b obtained from the final RV time-series analysis.}
        \label{tab:GJ740b_Properties}
        \begin{tabular}{lcc}
                \hline
                Parameter                 &  Priors  &  Value                                               \\
                \hline
                \multicolumn{3}{c}{MCMC}                                                                    \\
            \hline
                $K_{\rm b}$ [m\,s$^{-1}$] & $\mathcal{U}$ (0.0, 5.0)      & 2.13$^{+0.34}_{-0.32}$          \\
                $P_{\rm b}$ [d]           & $\mathcal{U}$ (1.5, 3.5)      & 2.37756$^{+0.00013}_{-0.00011}$ \\
                $T_{\rm b}$-2454647.7 [d] & $\mathcal{U}$ (0.0, 1.6)      & 0.87$^{+0.32}_{-0.30}$          \\
                $\omega_{\rm b}$ [rad]    & $\mathcal{U}$ (-$\pi$, $\pi$) & $-$0.72$^{+1.02}_{-0.84}$       \\
                $e_{\rm b}$               & $\mathcal{U}$ (0.0, 1.0)      & 0.24$^{+0.15}_{-0.14}$          \\
                \hline
                \multicolumn{3}{c}{Derived}                                              \\
                \hline
                $M_{\rm p} \sin$ i [M$_{\oplus}$] & $\cdots$ & 2.96$^{+0.50}_{-0.48}$    \\
                $a$ [AU]                          & $\cdots$ & 0.029$^{+0.001}_{-0.001}$ \\
                $T_{\rm eq}$ [K] $^{(*)}$         & $\cdots$ & 829$^{+40}_{-50}$         \\
                Insolation [S$_{\oplus}$]         & $\cdots$ & 79$^{+16}_{-17}$          \\
                Transit Probability               & $\cdots$ & 9.0\%                     \\
                \hline
        \end{tabular}
        \begin{minipage}{\columnwidth}
        {\footnotesize \textbf{Note:} $^{(*)}$ Computed assuming null bond albedo.}
\end{minipage}  
\end{table}

Figure~\ref{fig:RV_MCMC_SineRot1Pkepler_fixRot_2} shows how the RV values are fitted nicely by the planetary model, leaving minor residuals. As shown in the bottom panel of Fig.~\ref{fig:RV_MCMC_SineRot1Pkepler_fixRot_2}, the RMS of the residuals after subtracting the final model is only 1\,m\,s$^{-1}$. The periodogram of these residuals produces only non-significant signals (i.e., lower than the PSD value related to the 10\%~level of FAP), which is similar to the distribution found in the other activity indices after subtracting the rotation and the long-term signal. We added both a Keplerian and a sinusoidal model to try to track the presence of an additional planetary signal in the RV residuals, but the simulations did not match the convergence criteria.  

\subsection{Photometric transits}

Taking into account the high transit probability shown in Table~\ref{tab:GJ740b_Properties}, we computed the BoxLeastSquares (BLS) periodogram \citep{2002A&A...391..369K} for the ASAS time series and carried out an MCMC analysis similar to the one performed in the RV time series (without modeling any long-term signal since it was not detected in photometry) to search for the planetary signal. The posterior distribution of the planetary parameters did not meet the convergence criteria and therefore there is no evidence of detection . The SuperWASP and EXORAP datasets do not show any hint of a short-period signal in the P~$<$~10~d region with enough statistical significance to be reliable.

\subsection{Exploring additional signals}

To ensure that the GP regression is not overfitting and absorbing signals that are not related to stellar activity, we replaced the GP rotation model with a simpler one based on a double sinusoidal function. The periodogram of the RV residuals after subtracting this new model reveals the presence of a previously non-detected 15\,d signal. The inclusion of an additional sinusoidal function in our MCMC model to fit this signal (along with the long-term sinusoidal, the GP rotation term, and the GJ~740~b Keplerian) provides a greater $\log Z$ value than the one related to the previous model (without the short-term sinusoidal). However, the amplitude of this signal is below the 3$\sigma$ significance level and its periodicity requires a narrow prior to be constrained. Additionally, this 15\,d signal is also present in the H${\rm \alpha}$ and S$_{\rm mw}$ indicators. We performed the same MCMC analysis on these time series, and we obtained a good convergence based on the auto-correlation of the chains, which indicates that this signal is most likely caused by stellar activity.

\section{Discussion}

\label{sec:Discussion}

\subsection{Stellar activity}

We compared our results with those reported in \citet{2018A&A...612A..89S}, where the long-term signal of GJ~740 was detected with a periodicity of 2040\,d in the FWHM, H${\rm \alpha}$, and S$_{\rm mw}$ time series with lower semi-amplitudes (2.58\,m\,s$^{-1}$, 0.165, and 0.00658, respectively). The differences with the results presented in this paper could be explained by the larger dataset used in our analysis. Although the baseline of our data is sufficient to have a good estimation of the period of the presumed magnetic cycle, the most significant long-term peak differs between the different time series. Merging the probability distribution of the FWHM, H${\rm \alpha}$, S$_{\rm mw}$, and NaD datasets, we obtain a mean value of 2800$\pm$150\,d.

In the analysis presented in \citet{2018A&A...612A..89S}, we also detected the rotation period at 38\,d in the H${\rm \alpha}$ and S$_{\rm mw}$ time series, with an additional peak in the first harmonic at~$\sim$~19\,d. In the case of the FWHM time series, the peaks were shifted to 35 and 18\,d, respectively. The presence of the first harmonic of rotation in these time series could be caused by the geometric distribution of active regions. The photometry analysis carried out in the article supported the detection of the rotation signal at 35\,d. The average results from our analysis including the NaD time series characterized the rotation of the star with a periodicity of 36.5$\pm$1.0\,d.

Regarding the photometric variability, the EXORAP results suggest a scenario where this variability is affected by the effects of an irregularly spotted stellar surface coupled with stellar rotation. This scenario is consistent with the fact that the activity signal is stronger at bluer wavelengths, where the contrast between the photosphere and cool spots is larger. Furthermore, we notice that $B$ and $V$ photometry get dimmer with time, suggesting that the spot coverage increases during the observation campaign. This is consistent with an increasing level of stellar activity as also suggested by the chromospheric indices shown in Fig.~\ref{fig:Compact_Indexes}.

The coherence time obtained in the posterior distribution of the rotation parameters in the EXORAP analysis is comparable with the stellar rotation, which contrasts with the results obtained in \citet{2017A&A...598A..28S}, where the evolutionary timescale of active regions found shows typically longer values, on the order of a few stellar rotations. This indicates that the photosphere of GJ~740 is more dynamic than what is typically found for field M-dwarfs. The amplitude of the correlated noise decreases with increasing wavelength. Consistently with the periodogram analysis, this suggests that the correlated signal is due to the presence of cool spots corotating with the stellar surface. The stellar dimming during the survey is confirmed, supporting the scenario where the spot coverage (and the activity level) increases with time.

\subsection{GJ 740 b}

\begin{figure}
        \includegraphics[width=\columnwidth]{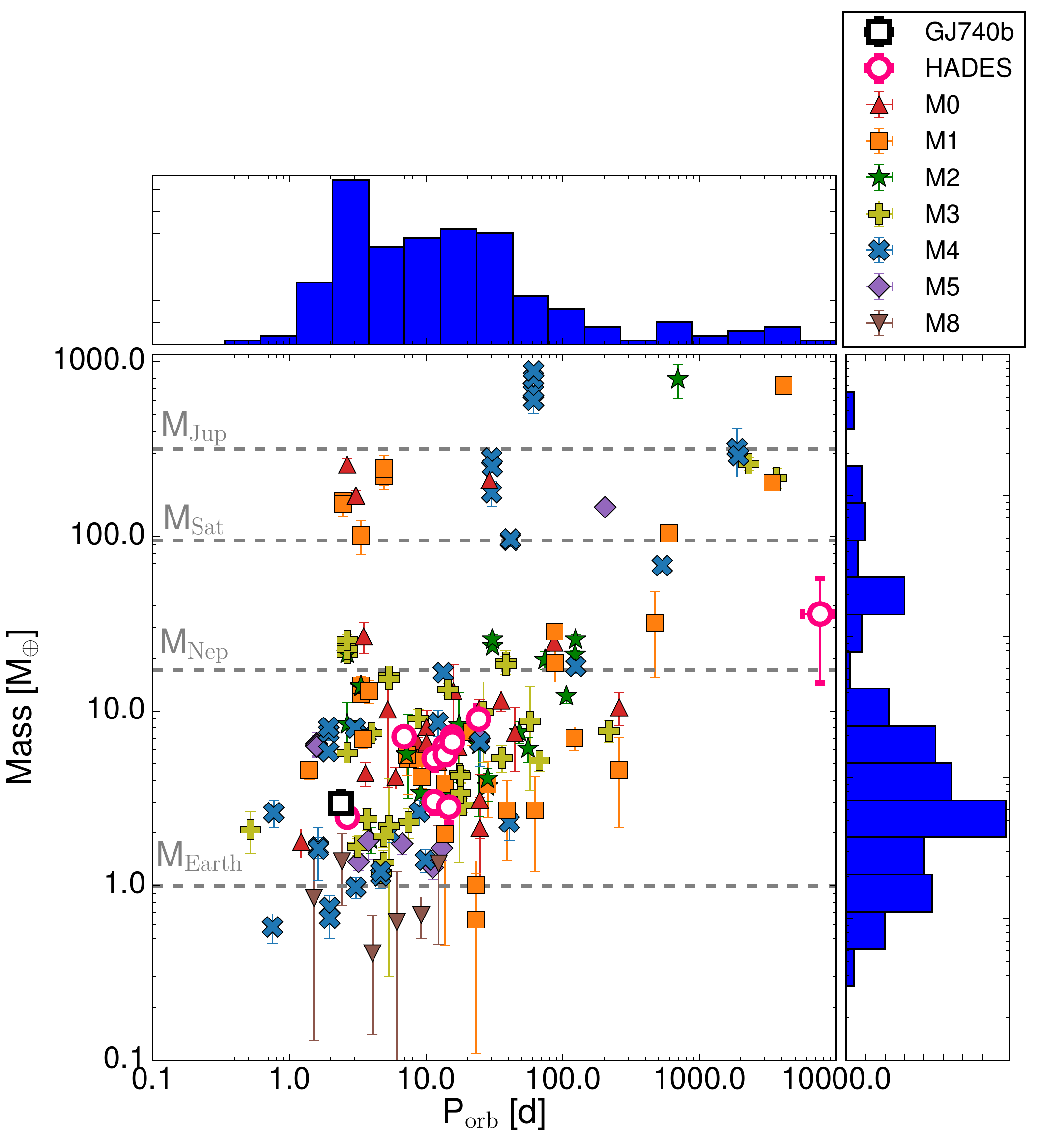}
    \caption{Mass-period diagram including the detected exoplanets from NASA exoplanets archive with published masses and orbital periods orbiting around M-type stars. The sub-spectral type of the parent star is indicated with a unique symbol and color. GJ~740~b has been represented with an unfilled black square, and the HADES detections (GJ~3998~b and GJ~3998~c, \citealt{2016A&A...593A.117A} ; GJ~625~b, \citealt{2017A&A...605A..92S} : GJ~3942~b, \citealt{2017A&A...608A..63P} ; Gl~15~A~b and Gl~15~A~c, \citealt{2018A&A...617A.104P} ; Gl~686~b, \citealt{2019A&A...622A.193A} ; Gl~49~b, \citealt{2019A&A...624A.123P} ; and GJ~685~b, \citealt{2019A&A...625A.126P}) have been marked with pink unfilled dots. The four horizontal dashed lines indicate the mass of Jupiter, Saturn, Neptune, and the Earth as a reference. The top and right panels contain the mass and orbital period distribution of the sample.}
    \label{fig:Diagrams_massVSporb}
\end{figure}

With regard to the planetary signal, the semi-major axis value shown in Table~\ref{tab:GJ740b_Properties} positions GJ~740~b in a close-in orbit to its parent star. We computed the habitable zone of GJ~740 based on the methodologies published by \citet{2007A&A...476.1373S} and \citet{2013ApJ...767L...8K}. The first one provides a range between 0.14 and 0.66\,AU, while the second one results in a conservative range between 0.25 and 0.48\,AU, and an optimistic range between 0.20 and 0.51\,AU. Therefore, we found that GJ~740~b is located out of the habitable zone of its parent star. The lack of a radius measurement does not allow for a precise description of  the composition of GJ~740~b with theoretical models, but its mass and short orbital period suggest a rocky composition \citep{2014ApJ...783L...6W}. The posterior distribution of the eccentricity of the planet presented in Fig.~\ref{fig:RV_MCMC_SineRot1Pkepler_fixRot_1} shows compatibility with a null value at 2$\sigma$. This has been proven to be usual for short-period Keplerian orbits \citep{2013MNRAS.434L..51K} and cases similar to the one of GJ~740~b can be found in the literature \citep{2017A&A...602A..88A,2017A&A...597A..34M,2019ApJS..242...25F}. To compare GJ~740~b with other detected planets around M-dwarfs with measured masses we created the mass-period diagram represented in Fig.~\ref{fig:Diagrams_massVSporb}.

Figure~\ref{fig:Diagrams_massVSporb} shows that GJ~740~b falls within the peak of the orbital period distribution and is very close to the mass peak. It is positioned as the planet with the second shortest orbital period around an M1 star after L 168-9 b \citep{2020A&A...636A..58A}. The super-Earth region of the diagram (between 2 and 10\,M$_{\oplus}$) is the most crowded zone, lacking any detected sub-Neptunes, Neptune-like, and Jovian planets at short periods. The diagram also shows a gap of low-mass planets with long periods due to instrument limitations. 

The search for a photometric counterpart of the planetary signal caused by the transits of the planet did not reveal any match within our photometric datasets. Using the mass-radius relation for exoplanets found by \citet{2020A&A...634A..43O}, we estimate a radius of 1.43$^{+0.12}_{-0.11}$\,R$_{\oplus}$ for GJ~740~b, assuming a density of $\rho_{\rm P} >$~3.3\,g\,cm$^{-3}$. Such a radius value leads to a transit depth of 0.5\,mmag, which is out of the precision range provided by the photometric instruments used in this work. The lack of a \textit{TESS} light curve for this target precludes a deeper analysis of this possible photometric signal. Future \textit{TESS} observations on this target are planned between 9 July and 5 August of 2022 within Sector 54. The \textit{CHEOPS} telescope would be an ideal instrument to check for the occurrence of transits. 

\subsection{Possible second planetary companion}

\begin{figure}
        \includegraphics[width=\columnwidth]{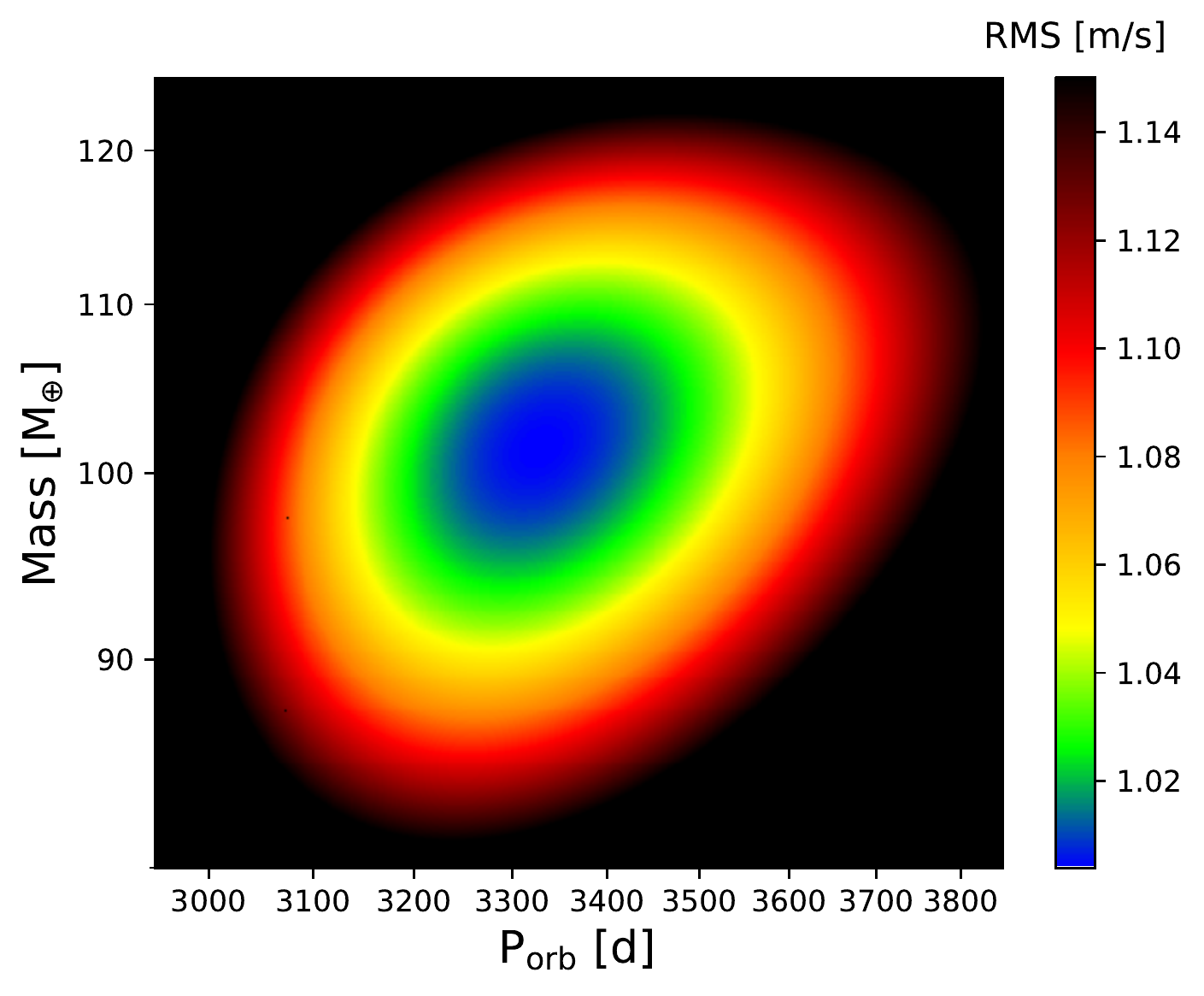}
    \caption{Distribution of the RMS of the residuals in the RV time series after subtracting the long-term signal with a sinusoidal function whose amplitude is calculated from the considered values of mass and period.}
    \label{fig:SecondPlanet_ProbabilityDistribution}
\end{figure}

We studied the possibility of having a second planet causing the~$\sim$~3400\,d signal in the RV time series since it is not clear that the origin of the signal is related to the presumed activity cycle of the star due to the differences with respect to the results obtained from the activity indicators. We calculated the range of masses associated with the plausible orbital period of this planet in Fig.~\ref{fig:SecondPlanet_ProbabilityDistribution} using a sinusoidal model.

Figure~\ref{fig:SecondPlanet_ProbabilityDistribution} indicates that this second planet would be characterized by a mass of~$\sim$~100\,M$_{\oplus}$. The existence of this high-mass companion is favored by the greater statistical possibility of finding super-Earths like GJ~740~b in multi-planetary systems \citep{2013ApJS..204...24B,2018Natur.563..365R}. Nevertheless, super-Earths with short orbital periods have been proven to be more likely to be on single-planetary systems than their analogs with longer orbital periods \citep{2018AJ....156..254W}; although only a few giant planets have been detected around M-dwarfs \citep{2010A&A...511A..21C,2011A&A...526A.141F,2012ApJ...746...37A,2013A&A...549A...2L,2019Sci...365.1441M}. In Fig.~\ref{fig:Diagrams_massVSporb}, only two planets with a period greater than 1000\,d orbiting a M1 star are shown: GJ~328~b \citep{2013ApJ...774..147R} and GJ~832~b \citep{2009ApJ...690..743B}. Both of them are Jupiter like planets with a mass of 2.30$\pm$0.13\,M$_{\rm Jup}$ and 0.64$\pm$0.06\,M$_{\rm Jup}$, and orbital separation of 4.5$\pm$0.2\,AU and 3.4$\pm$0.4\,AU, respectively. Implementing a Keplerian model to trace the signal results in a not well-defined eccentricity for the model due to the long period of the signal in comparison with the time span of the observations. 

To explore the possible origins of this signal, we performed a correlation study between the RV and the stellar activity indicators based on the Pearson coefficient \citep{1895RSPS...58..240P}. We calculated this coefficient along with the p-value crossing all the time series for each individual spectrograph first, obtaining a low non-significant correlation in HARPS and an intermediate significant correlation in HARPS-N. This indicates a different behavior in the stellar activity in the epoch when the HARPS-N measurements were taken. We then subtract the S$_{\rm mw}$ contribution to the RV time series, which causes a decrease in the PSD associated with the long-term signal in the periodogram but keeping a FAP below the 0.1\% level. Thus, we cannot conclude that this signal is entirely related to the stellar activity of GJ~740.

\begin{figure}
        \includegraphics[width=\columnwidth]{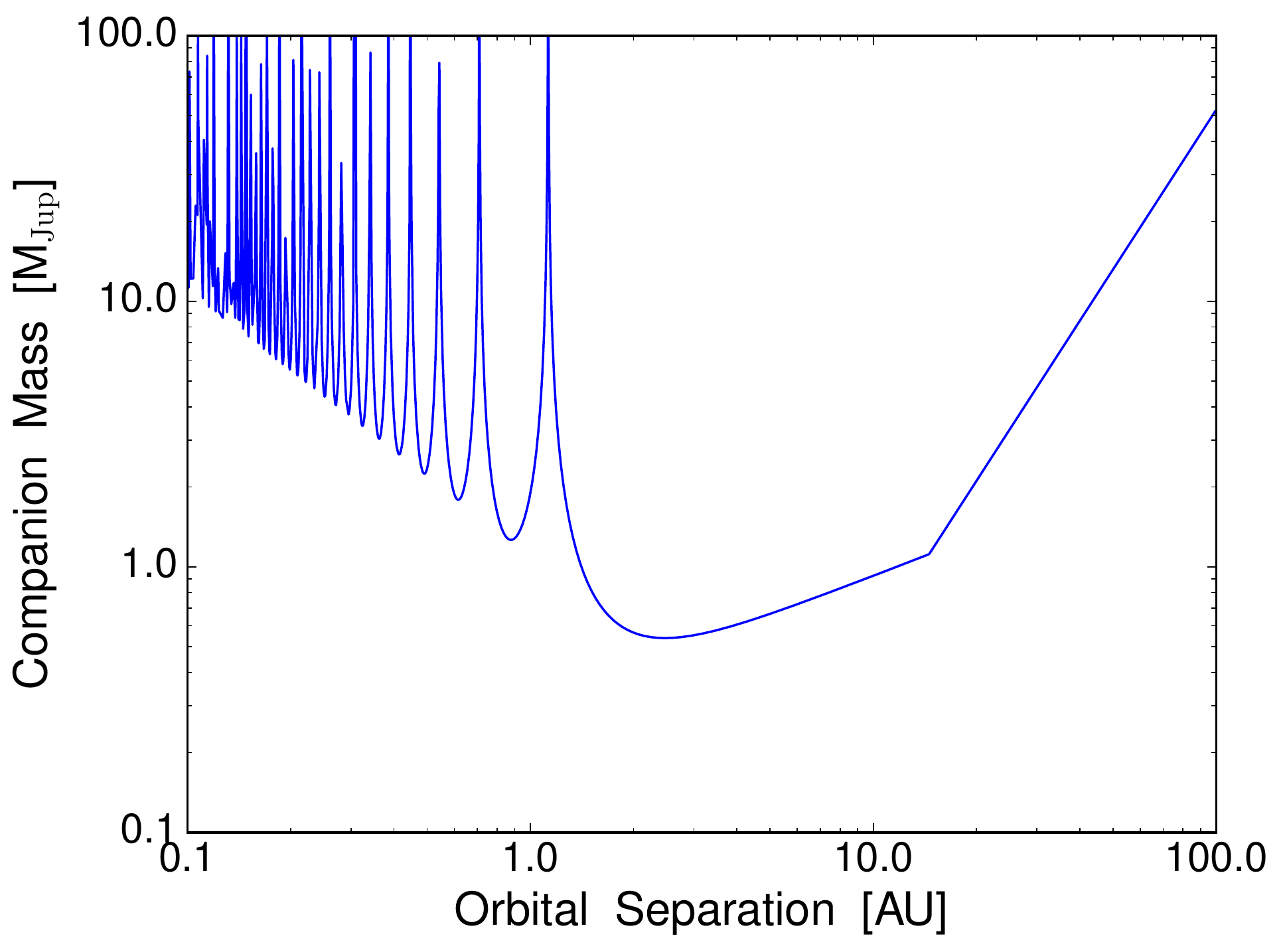}
    \caption{Diagram of the minimum mass of a planetary companion for GJ~740 at different orbital radius based on the proper motion difference method.}
    \label{fig:G740_PMA_sensitivity}
\end{figure}

Considering that GJ~740 is sufficiently bright to be observed by both Gaia and Hipparcos, we quantified the detection limits in the mass-separation diagram based on the proper motion difference technique. Using the formalism presented in \citet{2019A&A...623A..72K} (Equations 13, 14, and 15) we produced the diagram shown in Fig.~\ref{fig:G740_PMA_sensitivity}.

Although there is no evidence for a statistically significant proper motion anomaly in GJ~740, Fig.~\ref{fig:G740_PMA_sensitivity} indicates that an object at 3-4 AU (encompassing the orbital period of 9.3\,yr of the candidate planet) with a mass of around 0.6\,M$_{\rm Jup}$ can be ruled out at the 1-sigma level. This means that we can place an approximate limit on the inclination of the possible companion around 30\,deg.  To acquire sensitivity to a Saturn-mass object, such as the one we might be seeing in the RV time series, we will have to wait for future Gaia data release, starting with DR3 (in less than two years' time). An improvement in mass sensitivity of a factor of 2 at that period is likely to be achieved by combining improved calibration schemes for bright stars such as GJ~740 and more data undergoing processing.

The possibility that the signal is related to the magnetic cycle of the star is not clear due to the uncertainties in the periodicity of the signal from all the time series and the time coverage of our dataset, which is not enough to trace two periods of the cycle. This explanation for the signal is difficult to prove if we look at the low number of M-dwarfs with published and well-measured long-period cycles in the literature \citep{2012A&A...541A...9G,2012ARep...56..716S,2013ApJ...764....3R,2019MNRAS.488.5145T}, along with the probability that an M-dwarf that presents long-term activity variability may not present RV changes related to the magnetic cycle \citep{2012A&A...541A...9G}. The peak of the cycle length for early M-type stars has been located around 6\,yr \citep{2012A&A...541A...9G, 2016A&A...595A..12S}. The case of BD-114672, studied by \citet{2020A&A...641A..68B}, is a similar case to GJ~740, showing how a late K-type star with a mass similar to GJ~740 can exhibit both a long-period planet and cycle. Consequently, we conclude that the origin of the long-term RV signal is unclear until further observations are carried out on GJ~740.

Considering the mass of the candidate, the moderately old age of the system, and a favorable projected separation of~$\sim$~0.3", this target is still out of reach for current instrumentation such as SPHERE or GPI -- unless the planet is much brighter than expected (e.g., the possible detection of Proxima~Centauri~c with SPHERE presented in \citealt{2020A&A...638A.120G}). The new generation instruments at 30-40\,m class telescopes presents much better perspectives for this kind of detection.

\section{Conclusions}

\label{sec:Conclusions}

Our analysis of the 129 HARPS-N, 57 HARPS, and 32 CARMENES spectra of GJ~740 taken over 11\,yr shows the presence of a super-Earth orbiting the star with an orbital period of 2.37756$^{+0.00013}_{-0.00011}$\,d and an RV semi-amplitude of 2.13$^{+0.34}_{-0.32}$\,m\,s$^{-1}$. This planet presents a minimum mass of $M_{\rm p}\sin i$=2.96$^{+0.50}_{-0.48}$\,M$_{\oplus}$ and a transit probability of 9\%. We analyzed 474 photometric measurements from ASAS, 2350 SuperWASP measurements, and 5\,years' worth of EXORAP measurements to search for a possible periodic signal caused by the transit of the planet in these time series, however, none of the peaks in the periodogram around the orbital period of the planet present enough statistical significance.

Our study of five different spectroscopic time series reveals that GJ~740 presents variations consistent with a long-term cycle of 7.67$\pm$0.41\,yr and a rotation period of 36.5$\pm$1.0\,d. The photometric dataset of ASAS and EXORAP confirms the rotation of the star at 35.60$^{+0.89}_{-0.55}$\,d and 35.563~$\pm$0.071\,d, respectively. The RV time-series presents hints of a possible second planetary signal at 9.3\,yr that is also compatible with the presumed cycle signal detected in the stellar indices at 2$\sigma$. The MCMC analysis carried out on this signal did not converge to a clear eccentricity value and its origin cannot be determined given the time span of our current dataset.

\begin{acknowledgements}

B.T.P. acknowledges Fundaci\'on La Caixa for the financial support received in the form of a Ph.D. contract. A.S.M. acknowledges financial support from the Spanish Ministry of Science and Innovation (MICINN) under the 2019 Juan de la Cierva Programme. J.I.G.H. acknowledges financial support from Spanish MICINN under the 2013 Ram\'on y Cajal program RYC-2013-14875. B.T.P., A.S.M., J.I.G.H., R.R. acknowledge financial support from the Spanish MICINN AYA2017-86389-P. 

I.R. and M.Pe. acknowledge support from the Spanish MICINN and the Fondo Europeo de Desarrollo Regional (FEDER) through grant PGC2018-098153-B-C33, as well as the support of the Generalitat de Catalunya/CERCA program. GAPS acknowledges support from INAF through the Progetti Premiali funding scheme of the Italian Ministry of Education, University, and Research. GAPS acknowledges financial support from Progetto Premiale 2015 FRONTIERA (OB.FU. 1.05.06.11) funding scheme of the Italian Ministry of Education, University, and Research. G.S. acknowledges the funding support from Italian Space Agency (ASI) regulated by “Accordo ASI-INAF n. 2013-016-R.0 del 9 luglio 2013 e integrazione del 9 luglio 2015”. This research has received financial support from the agreement ASI-INAF n.2018-16-HH.0. 

The results of this paper were based on observations made with the Italian Telescopio Nazionale Galileo (TNG), operated on the island of La Palma by the INAF-Fundaci\'on Galileo Galilei at the Roque de Los Muchachos Observatory of the Instituto de Astrof\'isica de Canarias (IAC); observations made with the HARPS instrument on the ESO 3.6-m telescope at La Silla Observatory (Chile); and observations made with the CARMENES instrument.

CARMENES is an instrument for the Centro Astronómico Hispano-Alemán de Calar Alto (CAHA, Almería, Spain). CARMENES is funded by the German Max-Planck-Gesellschaft (MPG), the Spanish Consejo Superior de Investigaciones Científicas (CSIC), the European Union through FEDER/ERF FICTS-2011-02 funds, and the members of the CARMENES Consortium (Max-Planck-Institut für Astronomie, Instituto de Astrofísica de Andalucía, Landessternwarte Königstuhl, Institut de Ciències de l’Espai, Insitut für Astrophysik Göttingen, Universidad Complutense de Madrid, Thüringer Landessternwarte Tautenburg, Instituto de Astrofísica de Canarias, Hamburger Sternwarte, Centro de Astrobiología and Centro Astronómico Hispano-Alemán), with additional contributions by the Spanish MICINN through projects RYC2013-14875, AYA2015-69350-C3-2-P, AYA2016-79425-C3-1/2/3-P, ESP2016-80435-C2-1-R, ESP2017-87143-R, ESP2017-87676-C05-1/2/5-R, and AYA2017-86389-P, the German Science Foundation through the Major Research Instrumentation Program and DFG Research Unit FOR2544 “Blue Planets around Red Stars”, the Klaus Tschira Stiftung, the states of Baden-Württemberg and Niedersachsen, and by the Junta de Andalucía.

This paper made use of the IAC Supercomputing facility HTCondor (\url{http://research.cs.wisc.edu/htcondor/}), partly financed by the MICINN with FEDER funds, code IACA13-3E-2493. This research has made use of the NASA Exoplanet Archive, which is operated by the California Institute of Technology, under contract with the National Aeronautics and Space Administration under the Exoplanet Exploration Program. This paper makes use of data from the first public release of the WASP data (Butters et al. 2010) as provided by the WASP consortium and services at the NASA Exoplanet Archive, which is operated by the California Institute of Technology, under contract with the National Aeronautics and Space Administration under the Exoplanet Exploration Program. This work has made use of data from the European Space Agency (ESA) mission {\textit{Gaia}} (\url{https://www.cosmos.esa.int/gaia}), processed by the {\textit{Gaia}} Data Processing and Analysis Consortium (DPAC, \url{https://www.cosmos.esa.int/web/gaia/dpac/consortium}). Funding for the DPAC has been provided by national institutions, in particular, the institutions participating in the {\textit{Gaia}} Multilateral Agreement.

\end{acknowledgements}

%
   \bibliographystyle{aa} 
   \bibliography{References} 

\begin{thebibliography}{114}
\expandafter\ifx\csname natexlab\endcsname\relax\def\natexlab#1{#1}\fi

\bibitem[{{Affer} {et~al.}(2019){Affer}, {Damasso}, {Micela}, {Poretti},
  {Scandariato}, {Maldonado}, {Lanza}, {Covino}, {Garrido Rubio}, {Gonz{\'a}lez
  Hern{\'a}ndez}, {Gratton}, {Leto}, {Maggio}, {Perger}, {Sozzetti},
  {Su{\'a}rez Mascare{\~n}o}, {Bonomo}, {Borsa}, {Claudi}, {Cosentino},
  {Desidera}, {Giacobbe}, {Molinari}, {Pedani}, {Pinamonti}, {Rebolo}, {Ribas},
  \& {Toledo-Padr{\'o}n}}]{2019A&A...622A.193A}
{Affer}, L., {Damasso}, M., {Micela}, G., {et~al.} 2019, \aap, 622, A193

\bibitem[{{Affer} {et~al.}(2016){Affer}, {Micela}, {Damasso}, {Perger},
  {Ribas}, {Su{\'a}rez Mascare{\~n}o}, {Gonz{\'a}lez Hern{\'a}ndez}, {Rebolo},
  {Poretti}, {Maldonado}, {Leto}, {Pagano}, {Scandariato}, {Zanmar Sanchez},
  {Sozzetti}, {Bonomo}, {Malavolta}, {Morales}, {Rosich}, {Bignamini},
  {Gratton}, {Velasco}, {Cenadelli}, {Claudi}, {Cosentino}, {Desidera},
  {Giacobbe}, {Herrero}, {Lafarga}, {Lanza}, {Molinari}, \&
  {Piotto}}]{2016A&A...593A.117A}
{Affer}, L., {Micela}, G., {Damasso}, M., {et~al.} 2016, \aap, 593, A117

\bibitem[{{Ambikasaran} {et~al.}(2015){Ambikasaran}, {Foreman-Mackey},
  {Greengard}, {Hogg}, \& {O'Neil}}]{2015ITPAM..38..252A}
{Ambikasaran}, S., {Foreman-Mackey}, D., {Greengard}, L., {Hogg}, D.~W., \&
  {O'Neil}, M. 2015, IEEE Transactions on Pattern Analysis and Machine
  Intelligence, 38, 252

\bibitem[{{Anglada-Escud{\'e}} {et~al.}(2016){Anglada-Escud{\'e}}, {Amado},
  {Barnes}, {Berdi{\~n}as}, {Butler}, {Coleman}, {de La Cueva}, {Dreizler},
  {Endl}, {Giesers}, {Jeffers}, {Jenkins}, {Jones}, {Kiraga}, {K{\"u}rster},
  {L{\'o}pez-Gonz{\'a}lez}, {Marvin}, {Morales}, {Morin}, {Nelson}, {Ortiz},
  {Ofir}, {Paardekooper}, {Reiners}, {Rodr{\'\i}guez},
  {Rodr{\'\i}guez-L{\'o}pez}, {Sarmiento}, {Strachan}, {Tsapras}, {Tuomi}, \&
  {Zechmeister}}]{2016Natur.536..437A}
{Anglada-Escud{\'e}}, G., {Amado}, P.~J., {Barnes}, J., {et~al.} 2016, \nat,
  536, 437

\bibitem[{{Anglada-Escud{\'e}} {et~al.}(2012){Anglada-Escud{\'e}}, {Boss},
  {Weinberger}, {Thompson}, {Butler}, {Vogt}, \&
  {Rivera}}]{2012ApJ...746...37A}
{Anglada-Escud{\'e}}, G., {Boss}, A.~P., {Weinberger}, A.~J., {et~al.} 2012,
  \apj, 746, 37

\bibitem[{{Anglada-Escud{\'e}} \& {Butler}(2012)}]{2012ApJS..200...15A}
{Anglada-Escud{\'e}}, G. \& {Butler}, R.~P. 2012, \apjs, 200, 15

\bibitem[{{Astudillo-Defru} {et~al.}(2020){Astudillo-Defru}, {Cloutier},
  {Wang}, {Teske}, {Brahm}, {Hellier}, {Ricker}, {Vand erspek}, {Latham},
  {Seager}, {Winn}, {Jenkins}, {Collins}, {Stassun}, {Ziegler}, {Almenara},
  {Anderson}, {Artigau}, {Bonfils}, {Bouchy}, {Brice{\~n}o}, {Butler},
  {Charbonneau}, {Conti}, {Crane}, {Crossfield}, {Davies}, {Delfosse},
  {D{\'\i}az}, {Doyon}, {Dragomir}, {Eastman}, {Espinoza}, {Essack}, {Feng},
  {Figueira}, {Forveille}, {Gan}, {Glidden}, {Guerrero}, {Hart}, {Henning},
  {Horch}, {Isopi}, {Jenkins}, {Jord{\'a}n}, {Kielkopf}, {Law}, {Lovis},
  {Mallia}, {Mann}, {de Medeiros}, {Melo}, {Mennickent}, {Mignon}, {Murgas},
  {Nusdeo}, {Pepe}, {Relles}, {Rose}, {Santos}, {S{\'e}gransan}, {Shectman},
  {Shporer}, {Smith}, {Torres}, {Udry}, {Villasenor}, {Winters}, \&
  {Zhou}}]{2020A&A...636A..58A}
{Astudillo-Defru}, N., {Cloutier}, R., {Wang}, S.~X., {et~al.} 2020, \aap, 636,
  A58

\bibitem[{{Astudillo-Defru} {et~al.}(2017{\natexlab{a}}){Astudillo-Defru},
  {Delfosse}, {Bonfils}, {Forveille}, {Lovis}, \&
  {Rameau}}]{2017A&A...600A..13A}
{Astudillo-Defru}, N., {Delfosse}, X., {Bonfils}, X., {et~al.}
  2017{\natexlab{a}}, \aap, 600, A13

\bibitem[{{Astudillo-Defru} {et~al.}(2017{\natexlab{b}}){Astudillo-Defru},
  {D{\'\i}az}, {Bonfils}, {Almenara}, {Delisle}, {Bouchy}, {Delfosse},
  {Forveille}, {Lovis}, {Mayor}, {Murgas}, {Pepe}, {Santos}, {S{\'e}gransan},
  {Udry}, \& {W{\"u}nsche}}]{2017A&A...605L..11A}
{Astudillo-Defru}, N., {D{\'\i}az}, R.~F., {Bonfils}, X., {et~al.}
  2017{\natexlab{b}}, \aap, 605, L11

\bibitem[{{Astudillo-Defru} {et~al.}(2017{\natexlab{c}}){Astudillo-Defru},
  {Forveille}, {Bonfils}, {S{\'e}gransan}, {Bouchy}, {Delfosse}, {Lovis},
  {Mayor}, {Murgas}, {Pepe}, {Santos}, {Udry}, \&
  {W{\"u}nsche}}]{2017A&A...602A..88A}
{Astudillo-Defru}, N., {Forveille}, T., {Bonfils}, X., {et~al.}
  2017{\natexlab{c}}, \aap, 602, A88

\bibitem[{{Bailer-Jones} {et~al.}(2018){Bailer-Jones}, {Rybizki}, {Fouesneau},
  {Mantelet}, \& {Andrae}}]{2018AJ....156...58B}
{Bailer-Jones}, C.~A.~L., {Rybizki}, J., {Fouesneau}, M., {Mantelet}, G., \&
  {Andrae}, R. 2018, \aj, 156, 58

\bibitem[{{Bailey} {et~al.}(2009){Bailey}, {Butler}, {Tinney}, {Jones},
  {O'Toole}, {Carter}, \& {Marcy}}]{2009ApJ...690..743B}
{Bailey}, J., {Butler}, R.~P., {Tinney}, C.~G., {et~al.} 2009, \apj, 690, 743

\bibitem[{{Barbato} {et~al.}(2020){Barbato}, {Pinamonti}, {Sozzetti}, {Biazzo},
  {Benatti}, {Damasso}, {Desidera}, {Lanza}, {Maldonado}, {Mancini},
  {Scandariato}, {Affer}, {Andreuzzi}, {Bignamini}, {Bonomo}, {Borsa},
  {Carleo}, {Claudi}, {Cosentino}, {Covino}, {Fiorenzano}, {Giacobbe},
  {Harutyunyan}, {Knapic}, {Leto}, {Lorenzi}, {Maggio}, {Malavolta}, {Micela},
  {Molinari}, {Molinaro}, {Nascimbeni}, {Pagano}, {Pedani}, {Piotto},
  {Poretti}, \& {Rainer}}]{2020A&A...641A..68B}
{Barbato}, D., {Pinamonti}, M., {Sozzetti}, A., {et~al.} 2020, \aap, 641, A68

\bibitem[{{Batalha} {et~al.}(2013){Batalha}, {Rowe}, {Bryson}, {Barclay},
  {Burke}, {Caldwell}, {Christiansen}, {Mullally}, {Thompson}, {Brown},
  {Dupree}, {Fabrycky}, {Ford}, {Fortney}, {Gilliland}, {Isaacson}, {Latham},
  {Marcy}, {Quinn}, {Ragozzine}, {Shporer}, {Borucki}, {Ciardi}, {Gautier},
  {Haas}, {Jenkins}, {Koch}, {Lissauer}, {Rapin}, {Basri}, {Boss}, {Buchhave},
  {Carter}, {Charbonneau}, {Christensen-Dalsgaard}, {Clarke}, {Cochran},
  {Demory}, {Desert}, {Devore}, {Doyle}, {Esquerdo}, {Everett}, {Fressin},
  {Geary}, {Girouard}, {Gould}, {Hall}, {Holman}, {Howard}, {Howell},
  {Ibrahim}, {Kinemuchi}, {Kjeldsen}, {Klaus}, {Li}, {Lucas}, {Meibom},
  {Morris}, {Pr{\v{s}}a}, {Quintana}, {Sanderfer}, {Sasselov}, {Seader},
  {Smith}, {Steffen}, {Still}, {Stumpe}, {Tarter}, {Tenenbaum}, {Torres},
  {Twicken}, {Uddin}, {Van Cleve}, {Walkowicz}, \&
  {Welsh}}]{2013ApJS..204...24B}
{Batalha}, N.~M., {Rowe}, J.~F., {Bryson}, S.~T., {et~al.} 2013, \apjs, 204, 24

\bibitem[{{Bauer} {et~al.}(2015){Bauer}, {Zechmeister}, \&
  {Reiners}}]{2015A&A...581A.117B}
{Bauer}, F.~F., {Zechmeister}, M., \& {Reiners}, A. 2015, \aap, 581, A117

\bibitem[{{Berdi{\~n}as} {et~al.}(2017){Berdi{\~n}as},
  {Rodr{\'\i}guez-L{\'o}pez}, {Amado}, {Anglada-Escud{\'e}}, {Barnes},
  {MacDonald}, {Zechmeister}, \& {Sarmiento}}]{2017MNRAS.469.4268B}
{Berdi{\~n}as}, Z.~M., {Rodr{\'\i}guez-L{\'o}pez}, C., {Amado}, P.~J., {et~al.}
  2017, \mnras, 469, 4268

\bibitem[{{Bonfils} {et~al.}(2018){Bonfils}, {Astudillo-Defru}, {D{\'\i}az},
  {Almenara}, {Forveille}, {Bouchy}, {Delfosse}, {Lovis}, {Mayor}, {Murgas},
  {Pepe}, {Santos}, {S{\'e}gransan}, {Udry}, \&
  {W{\"u}nsche}}]{2018A&A...613A..25B}
{Bonfils}, X., {Astudillo-Defru}, N., {D{\'\i}az}, R., {et~al.} 2018, \aap,
  613, A25

\bibitem[{{Bonfils} {et~al.}(2007){Bonfils}, {Mayor}, {Delfosse}, {Forveille},
  {Gillon}, {Perrier}, {Udry}, {Bouchy}, {Lovis}, {Pepe}, {Queloz}, {Santos},
  \& {Bertaux}}]{2007A&A...474..293B}
{Bonfils}, X., {Mayor}, M., {Delfosse}, X., {et~al.} 2007, \aap, 474, 293

\bibitem[{{Butler} {et~al.}(2006){Butler}, {Johnson}, {Marcy}, {Wright},
  {Vogt}, \& {Fischer}}]{2006PASP..118.1685B}
{Butler}, R.~P., {Johnson}, J.~A., {Marcy}, G.~W., {et~al.} 2006, \pasp, 118,
  1685

\bibitem[{{Caballero} {et~al.}(2016){Caballero}, {Gu{\`a}rdia}, {L{\'o}pez del
  Fresno}, {Zechmeister}, {de Juan}, {Alonso-Floriano}, {Amado}, {Colom{\'e}},
  {Cort{\'e}s-Contreras}, {Garc{\'\i}a-Piquer}, {Gesa}, {de Guindos}, {Hagen},
  {Helmling}, {Hern{\'a}ndez Casta{\~n}o}, {K{\"u}rster}, {L{\'o}pez-Santiago},
  {Montes}, {Morales Mu{\~n}oz}, {Pavlov}, {Quirrenbach}, {Reiners}, {Ribas},
  {Seifert}, \& {Solano}}]{2016SPIE.9910E..0EC}
{Caballero}, J.~A., {Gu{\`a}rdia}, J., {L{\'o}pez del Fresno}, M., {et~al.}
  2016, in Society of Photo-Optical Instrumentation Engineers (SPIE) Conference
  Series, Vol. 9910, Observatory Operations: Strategies, Processes, and Systems
  VI, ed. A.~B. {Peck}, R.~L. {Seaman}, \& C.~R. {Benn}, 99100E

\bibitem[{{Carson} {et~al.}(2005){Carson}, {Eikenberry}, {Brandl}, {Wilson}, \&
  {Hayward}}]{2005AJ....130.1212C}
{Carson}, J.~C., {Eikenberry}, S.~S., {Brandl}, B.~R., {Wilson}, J.~C., \&
  {Hayward}, T.~L. 2005, \aj, 130, 1212

\bibitem[{{Chabrier} \& {Baraffe}(2000)}]{2000ARA&A..38..337C}
{Chabrier}, G. \& {Baraffe}, I. 2000, \araa, 38, 337

\bibitem[{{Cifuentes} {et~al.}(2020){Cifuentes}, {Caballero},
  {Cort{\'e}s-Contreras}, {Montes}, {Abell{\'a}n}, {Dorda}, {Holgado},
  {Zapatero Osorio}, {Morales}, {Amado}, {Passegger}, {Quirrenbach}, {Reiners},
  {Ribas}, {Sanz-Forcada}, {Schweitzer}, {Seifert}, \&
  {Solano}}]{2020A&A...642A.115C}
{Cifuentes}, C., {Caballero}, J.~A., {Cort{\'e}s-Contreras}, M., {et~al.} 2020,
  \aap, 642, A115

\bibitem[{{Correia} {et~al.}(2010){Correia}, {Couetdic}, {Laskar}, {Bonfils},
  {Mayor}, {Bertaux}, {Bouchy}, {Delfosse}, {Forveille}, {Lovis}, {Pepe},
  {Perrier}, {Queloz}, \& {Udry}}]{2010A&A...511A..21C}
{Correia}, A.~C.~M., {Couetdic}, J., {Laskar}, J., {et~al.} 2010, \aap, 511,
  A21

\bibitem[{{Cosentino} {et~al.}(2012){Cosentino}, {Lovis}, {Pepe}, {Collier
  Cameron}, {Latham}, {Molinari}, {Udry}, {Bezawada}, {Black}, {Born},
  {Buchschacher}, {Charbonneau}, {Figueira}, {Fleury}, {Galli}, {Gallie},
  {Gao}, {Ghedina}, {Gonzalez}, {Gonzalez}, {Guerra}, {Henry}, {Horne},
  {Hughes}, {Kelly}, {Lodi}, {Lunney}, {Maire}, {Mayor}, {Micela}, {Ordway},
  {Peacock}, {Phillips}, {Piotto}, {Pollacco}, {Queloz}, {Rice}, {Riverol},
  {Riverol}, {San Juan}, {Sasselov}, {Segransan}, {Sozzetti}, {Sosnowska},
  {Stobie}, {Szentgyorgyi}, {Vick}, \& {Weber}}]{2012SPIE.8446E..1VC}
{Cosentino}, R., {Lovis}, C., {Pepe}, F., {et~al.} 2012, in \procspie, Vol.
  8446, Ground-based and Airborne Instrumentation for Astronomy IV, 84461V

\bibitem[{{Covino} {et~al.}(2013){Covino}, {Esposito}, {Barbieri}, {Mancini},
  {Nascimbeni}, {Claudi}, {Desidera}, {Gratton}, {Lanza}, {Sozzetti}, {Biazzo},
  {Affer}, {Gandolfi}, {Munari}, {Pagano}, {Bonomo}, {Collier Cameron},
  {H{\'e}brard}, {Maggio}, {Messina}, {Micela}, {Molinari}, {Pepe}, {Piotto},
  {Ribas}, {Santos}, {Southworth}, {Shkolnik}, {Triaud}, {Bedin}, {Benatti},
  {Boccato}, {Bonavita}, {Borsa}, {Borsato}, {Brown}, {Carolo}, {Ciceri},
  {Cosentino}, {Damasso}, {Faedi}, {Mart{\'\i}nez Fiorenzano}, {Latham},
  {Lovis}, {Mordasini}, {Nikolov}, {Poretti}, {Rainer}, {Rebolo L{\'o}pez},
  {Scandariato}, {Silvotti}, {Smareglia}, {Alcal{\'a}}, {Cunial}, {Di
  Fabrizio}, {Di Mauro}, {Giacobbe}, {Granata}, {Harutyunyan}, {Knapic},
  {Lattanzi}, {Leto}, {Lodato}, {Malavolta}, {Marzari}, {Molinaro},
  {Nardiello}, {Pedani}, {Prisinzano}, \& {Turrini}}]{2013A&A...554A..28C}
{Covino}, E., {Esposito}, M., {Barbieri}, M., {et~al.} 2013, \aap, 554, A28

\bibitem[{{Delfosse} {et~al.}(1998){Delfosse}, {Forveille}, {Mayor}, {Perrier},
  {Naef}, \& {Queloz}}]{1998A&A...338L..67D}
{Delfosse}, X., {Forveille}, T., {Mayor}, M., {et~al.} 1998, \aap, 338, L67

\bibitem[{{D{\'{\i}}az} {et~al.}(2007){D{\'{\i}}az}, {Cincunegui}, \&
  {Mauas}}]{2007MNRAS.378.1007D}
{D{\'{\i}}az}, R.~F., {Cincunegui}, C., \& {Mauas}, P.~J.~D. 2007, \mnras, 378,
  1007

\bibitem[{{D{\'\i}ez Alonso} {et~al.}(2019){D{\'\i}ez Alonso}, {Caballero},
  {Montes}, {de Cos Juez}, {Dreizler}, {Dubois}, {Jeffers}, {Lalitha}, {Naves},
  {Reiners}, {Ribas}, {Vanaverbeke}, {Amado}, {B{\'e}jar},
  {Cort{\'e}s-Contreras}, {Herrero}, {Hidalgo}, {K{\"u}rster}, {Logie},
  {Quirrenbach}, {Rau}, {Seifert}, {Sch{\"o}fer}, \&
  {Tal-Or}}]{2019A&A...621A.126D}
{D{\'\i}ez Alonso}, E., {Caballero}, J.~A., {Montes}, D., {et~al.} 2019, \aap,
  621, A126

\bibitem[{{Dreizler} {et~al.}(2020){Dreizler}, {Jeffers}, {Rodr{\'\i}guez},
  {Zechmeister}, {Barnes}, {Haswell}, {Coleman}, {Lalitha}, {Hidalgo Soto},
  {Strachan}, {Hambsch}, {L{\'o}pez-Gonz{\'a}lez}, {Morales}, {Rodr{\'\i}guez
  L{\'o}pez}, {Berdi{\~n}as}, {Ribas}, {Pall{\'e}}, {Reiners}, \&
  {Anglada-Escud{\'e}}}]{2020MNRAS.493..536D}
{Dreizler}, S., {Jeffers}, S.~V., {Rodr{\'\i}guez}, E., {et~al.} 2020, \mnras,
  493, 536

\bibitem[{{Dressing} \& {Charbonneau}(2015)}]{2015ApJ...807...45D}
{Dressing}, C.~D. \& {Charbonneau}, D. 2015, \apj, 807, 45

\bibitem[{{Dumusque} {et~al.}(2011){Dumusque}, {Santos}, {Udry}, {Lovis}, \&
  {Bonfils}}]{2011IAUS..276..527D}
{Dumusque}, X., {Santos}, N.~C., {Udry}, S., {Lovis}, C., \& {Bonfils}, X.
  2011, in The Astrophysics of Planetary Systems: Formation, Structure, and
  Dynamical Evolution, ed. A.~{Sozzetti}, M.~G. {Lattanzi}, \& A.~P. {Boss},
  Vol. 276, 527--529

\bibitem[{{Eastman} {et~al.}(2010){Eastman}, {Siverd}, \&
  {Gaudi}}]{2010PASP..122..935E}
{Eastman}, J., {Siverd}, R., \& {Gaudi}, B.~S. 2010, \pasp, 122, 935

\bibitem[{{Endl} {et~al.}(2006){Endl}, {Cochran}, {K{\"u}rster}, {Paulson},
  {Wittenmyer}, {MacQueen}, \& {Tull}}]{2006ApJ...649..436E}
{Endl}, M., {Cochran}, W.~D., {K{\"u}rster}, M., {et~al.} 2006, \apj, 649, 436

\bibitem[{{Endl} {et~al.}(2001){Endl}, {K{\"u}rster}, {Els}, {Hatzes}, \&
  {Cochran}}]{2001A&A...374..675E}
{Endl}, M., {K{\"u}rster}, M., {Els}, S., {Hatzes}, A.~P., \& {Cochran}, W.~D.
  2001, \aap, 374, 675

\bibitem[{{Feng} {et~al.}(2019){Feng}, {Crane}, {Xuesong Wang}, {Teske},
  {Shectman}, {D{\'{\i}}az}, {Thompson}, {Jones}, \&
  {Butler}}]{2019ApJS..242...25F}
{Feng}, F., {Crane}, J.~D., {Xuesong Wang}, S., {et~al.} 2019, \apjs, 242, 25

\bibitem[{{Foreman-Mackey} {et~al.}(2017){Foreman-Mackey}, {Agol},
  {Ambikasaran}, \& {Angus}}]{2017AJ....154..220F}
{Foreman-Mackey}, D., {Agol}, E., {Ambikasaran}, S., \& {Angus}, R. 2017, \aj,
  154, 220

\bibitem[{{Foreman-Mackey} {et~al.}(2013){Foreman-Mackey}, {Hogg}, {Lang}, \&
  {Goodman}}]{2013PASP..125..306F}
{Foreman-Mackey}, D., {Hogg}, D.~W., {Lang}, D., \& {Goodman}, J. 2013, \pasp,
  125, 306

\bibitem[{{Forveille} {et~al.}(2011){Forveille}, {Bonfils}, {Lo Curto},
  {Delfosse}, {Udry}, {Bouchy}, {Lovis}, {Mayor}, {Moutou}, {Naef}, {Pepe},
  {Perrier}, {Queloz}, \& {Santos}}]{2011A&A...526A.141F}
{Forveille}, T., {Bonfils}, X., {Lo Curto}, G., {et~al.} 2011, \aap, 526, A141

\bibitem[{{Gaia Collaboration} {et~al.}(2018){Gaia Collaboration}, {Brown},
  {Vallenari}, {Prusti}, {de Bruijne}, {Babusiaux}, {Bailer-Jones}, {Biermann},
  {Evans}, {Eyer}, \& et~al.}]{2018A&A...616A...1G}
{Gaia Collaboration}, {Brown}, A.~G.~A., {Vallenari}, A., {et~al.} 2018, \aap,
  616, A1

\bibitem[{{Gaudi} {et~al.}(2008){Gaudi}, {Bennett}, {Udalski}, {Gould},
  {Christie}, {Maoz}, {Dong}, {McCormick}, {Szyma{\'n}ski}, {Tristram},
  {Nikolaev}, {Paczy{\'n}ski}, {Kubiak}, {Pietrzy{\'n}ski}, {Soszy{\'n}ski},
  {Szewczyk}, {Ulaczyk}, {Wyrzykowski}, {OGLE Collaboration}, {DePoy}, {Han},
  {Kaspi}, {Lee}, {Mallia}, {Natusch}, {Pogge}, {Park}, {{\ensuremath{\mu}}-Fun
  Collabortion}, {Abe}, {Bond}, {Botzler}, {Fukui}, {Hearnshaw}, {Itow},
  {Kamiya}, {Korpela}, {Kilmartin}, {Lin}, {Masuda}, {Matsubara}, {Motomura},
  {Muraki}, {Nakamura}, {Okumura}, {Ohnishi}, {Rattenbury}, {Sako}, {Saito},
  {Sato}, {Skuljan}, {Sullivan}, {Sumi}, {Sweatman}, {Yock}, {MOA
  Collaboration}, {Albrow}, {Allan}, {Beaulieu}, {Burgdorf}, {Cook},
  {Coutures}, {Dominik}, {Dieters}, {Fouqu{\'e}}, {Greenhill}, {Horne},
  {Steele}, {Tsapras}, {Planet Collaboration}, {RoboNet Collaborations},
  {Chaboyer}, {Crocker}, {Frank}, \& {Macintosh}}]{2008Sci...319..927G}
{Gaudi}, B.~S., {Bennett}, D.~P., {Udalski}, A., {et~al.} 2008, Science, 319,
  927

\bibitem[{{Giacobbe} {et~al.}(2020){Giacobbe}, {Benedetto}, {Damasso},
  {Sozzetti}, {Christille}, {Lattanzi}, {Calcidese}, {Carbognani}, {Barbato},
  {Pinamonti}, {Poggio}, {Lanza}, {Bernagozzi}, {Cenadelli}, {Lanteri}, \&
  {Bertolini}}]{2020MNRAS.491.5216G}
{Giacobbe}, P., {Benedetto}, M., {Damasso}, M., {et~al.} 2020, \mnras, 491,
  5216

\bibitem[{{Giampapa} {et~al.}(1989){Giampapa}, {Cram}, \&
  {Wild}}]{1989ApJ...345..536G}
{Giampapa}, M.~S., {Cram}, L.~E., \& {Wild}, W.~J. 1989, \apj, 345, 536

\bibitem[{{Gomes da Silva} {et~al.}(2011){Gomes da Silva}, {Santos}, {Bonfils},
  {Delfosse}, {Forveille}, \& {Udry}}]{2011A&A...534A..30G}
{Gomes da Silva}, J., {Santos}, N.~C., {Bonfils}, X., {et~al.} 2011, \aap, 534,
  A30

\bibitem[{{Gomes da Silva} {et~al.}(2012){Gomes da Silva}, {Santos}, {Bonfils},
  {Delfosse}, {Forveille}, {Udry}, {Dumusque}, \&
  {Lovis}}]{2012A&A...541A...9G}
{Gomes da Silva}, J., {Santos}, N.~C., {Bonfils}, X., {et~al.} 2012, \aap, 541,
  A9

\bibitem[{{Gonz{\'a}lez-{\'A}lvarez} {et~al.}(2019){Gonz{\'a}lez-{\'A}lvarez},
  {Micela}, {Maldonado}, {Affer}, {Maggio}, {Lanza}, {Covino}, {Benatti},
  {Bignamini}, {Cosentino}, {Damasso}, {Desidera}, {Gonz{\'a}lez
  Hern{\'a}ndez}, {Mart{\'{\i}}nez-Fiorenzano}, {Pagano}, {Perger}, {Piotto},
  {Pinamonti}, {Rainer}, {Rebolo}, {Ribas}, {Scandariato}, {Sozzetti},
  {Su{\'a}rez Mascare{\~n}o}, \& {Toledo-Padr{\'o}n}}]{2019A&A...624A..27G}
{Gonz{\'a}lez-{\'A}lvarez}, E., {Micela}, G., {Maldonado}, J., {et~al.} 2019,
  \aap, 624, A27

\bibitem[{{Gonz{\'a}lez-{\'A}lvarez} {et~al.}(2020){Gonz{\'a}lez-{\'A}lvarez},
  {Zapatero Osorio}, {Caballero}, {Sanz-Forcada}, {B{\'e}jar},
  {Gonz{\'a}lez-Cuesta}, {Dreizler}, {Bauer}, {Rodr{\'\i}guez}, {Tal-Or},
  {Zechmeister}, {Montes}, {L{\'o}pez-Gonz{\'a}lez}, {Ribas}, {Reiners},
  {Quirrenbach}, {Amado}, {Anglada-Escud{\'e}}, {Azzaro},
  {Cort{\'e}s-Contreras}, {Hatzes}, {Henning}, {Jeffers}, {Kaminski},
  {K{\"u}rster}, {Lafarga}, {Morales}, {Pall{\'e}}, {Perger}, \&
  {Schmitt}}]{2020A&A...637A..93G}
{Gonz{\'a}lez-{\'A}lvarez}, E., {Zapatero Osorio}, M.~R., {Caballero}, J.~A.,
  {et~al.} 2020, \aap, 637, A93

\bibitem[{{Grandjean} {et~al.}(2020){Grandjean}, {Lagrange}, {Keppler},
  {Meunier}, {Mignon}, {Borgniet}, {Chauvin}, {Desidera}, {Galland}, {Messina},
  {Sterzik}, {Pantoja}, {Rodet}, \& {Zicher}}]{2020A&A...633A..44G}
{Grandjean}, A., {Lagrange}, A.~M., {Keppler}, M., {et~al.} 2020, \aap, 633,
  A44

\bibitem[{{Gratton} {et~al.}(2020){Gratton}, {Zurlo}, {Le Coroller}, {Damasso},
  {Del Sordo}, {Langlois}, {Mesa}, {Milli}, {Chauvin}, {Desidera}, {Hagelberg},
  {Lagadec}, {Vigan}, {Boccaletti}, {Bonnefoy}, {Brandner}, {Brown},
  {Cantalloube}, {Delorme}, {D'Orazi}, {Feldt}, {Galicher}, {Henning},
  {Janson}, {Kervella}, {Lagrange}, {Lazzoni}, {Ligi}, {Maire}, {M{\'e}nard},
  {Meyer}, {Mugnier}, {Potier}, {Rickman}, {Rodet}, {Romero}, {Schmidt},
  {Sissa}, {Sozzetti}, {Szul{\'a}gyi}, {Wahhaj}, {Antichi}, {Fusco}, {Stadler},
  {Suarez}, \& {Wildi}}]{2020A&A...638A.120G}
{Gratton}, R., {Zurlo}, A., {Le Coroller}, H., {et~al.} 2020, \aap, 638, A120

\bibitem[{{Hawley} {et~al.}(2014){Hawley}, {Davenport}, {Kowalski},
  {Wisniewski}, {Hebb}, {Deitrick}, \& {Hilton}}]{2014ApJ...797..121H}
{Hawley}, S.~L., {Davenport}, J.~R.~A., {Kowalski}, A.~F., {et~al.} 2014, \apj,
  797, 121

\bibitem[{{Haywood} {et~al.}(2014){Haywood}, {Collier Cameron}, {Queloz},
  {Barros}, {Deleuil}, {Fares}, {Gillon}, {Lanza}, {Lovis}, {Moutou}, {Pepe},
  {Pollacco}, {Santerne}, {S{\'e}gransan}, \& {Unruh}}]{2014MNRAS.443.2517H}
{Haywood}, R.~D., {Collier Cameron}, A., {Queloz}, D., {et~al.} 2014, \mnras,
  443, 2517

\bibitem[{{Hobson} {et~al.}(2019){Hobson}, {Delfosse}, {Astudillo-Defru},
  {Boisse}, {D{\'\i}az}, {Bouchy}, {Bonfils}, {Forveille}, {Arnold},
  {Borgniet}, {Bourrier}, {Brugger}, {Cabrera Salazar}, {Courcol}, {Dalal},
  {Deleuil}, {Demangeon}, {Dumusque}, {Hara}, {H{\'e}brard}, {Kiefer}, {Lopez},
  {Mignon}, {Montagnier}, {Mousis}, {Moutou}, {Pepe}, {Rey}, {Santerne},
  {Santos}, {Stalport}, {S{\'e}gransan}, {Udry}, \&
  {Wilson}}]{2019A&A...625A..18H}
{Hobson}, M.~J., {Delfosse}, X., {Astudillo-Defru}, N., {et~al.} 2019, \aap,
  625, A18

\bibitem[{{Horne} \& {Baliunas}(1986)}]{1986ApJ...302..757H}
{Horne}, J.~H. \& {Baliunas}, S.~L. 1986, \apj, 302, 757

\bibitem[{{Houdebine} {et~al.}(2009){Houdebine}, {Stempels}, \&
  {Oliveira}}]{2009MNRAS.400..238H}
{Houdebine}, E.~R., {Stempels}, H.~C., \& {Oliveira}, J.~H. 2009, \mnras, 400,
  238

\bibitem[{{Howard} {et~al.}(2010){Howard}, {Johnson}, {Marcy}, {Fischer},
  {Wright}, {Bernat}, {Henry}, {Peek}, {Isaacson}, {Apps}, {Endl}, {Cochran},
  {Valenti}, {Anderson}, \& {Piskunov}}]{2010ApJ...721.1467H}
{Howard}, A.~W., {Johnson}, J.~A., {Marcy}, G.~W., {et~al.} 2010, \apj, 721,
  1467

\bibitem[{{Kervella} {et~al.}(2019){Kervella}, {Arenou}, {Mignard}, \&
  {Th{\'e}venin}}]{2019A&A...623A..72K}
{Kervella}, P., {Arenou}, F., {Mignard}, F., \& {Th{\'e}venin}, F. 2019, \aap,
  623, A72

\bibitem[{{Kipping}(2013)}]{2013MNRAS.434L..51K}
{Kipping}, D.~M. 2013, \mnras, 434, L51

\bibitem[{{Kopparapu}(2013)}]{2013ApJ...767L...8K}
{Kopparapu}, R.~K. 2013, \apjl, 767, L8

\bibitem[{{Kov{\'a}cs} {et~al.}(2002){Kov{\'a}cs}, {Zucker}, \&
  {Mazeh}}]{2002A&A...391..369K}
{Kov{\'a}cs}, G., {Zucker}, S., \& {Mazeh}, T. 2002, \aap, 391, 369

\bibitem[{{K{\"u}rster} {et~al.}(2003){K{\"u}rster}, {Endl}, {Rouesnel}, {Els},
  {Kaufer}, {Brillant}, {Hatzes}, {Saar}, \& {Cochran}}]{2003A&A...403.1077K}
{K{\"u}rster}, M., {Endl}, M., {Rouesnel}, F., {et~al.} 2003, \aap, 403, 1077

\bibitem[{{Lalitha} {et~al.}(2019){Lalitha}, {Baroch}, {Morales}, {Passegger},
  {Bauer}, {Cardona Guill{\'e}n}, {Dreizler}, {Oshagh}, {Reiners}, {Ribas},
  {Caballero}, {Quirrenbach}, {Amado}, {B{\'e}jar}, {Colom{\'e}},
  {Cort{\'e}s-Contreras}, {Galad{\'{\i}}-Enr{\'{\i}}quez},
  {Gonz{\'a}lez-Cuesta}, {Guenther}, {Hagen}, {Henning}, {Herrero}, {Husser},
  {Jeffers}, {Kaminski}, {K{\"u}rster}, {Lafarga}, {Lodieu},
  {L{\'o}pez-Gonz{\'a}lez}, {Montes}, {Perger}, {Rosich}, {Rodr{\'{\i}}guez},
  {Rodr{\'{\i}}guez-L{\'o}pez}, {Schmitt}, {Tal-Or}, \&
  {Zechmeister}}]{2019A&A...627A.116L}
{Lalitha}, S., {Baroch}, D., {Morales}, J.~C., {et~al.} 2019, \aap, 627, A116

\bibitem[{{Lamman} {et~al.}(2020){Lamman}, {Baranec}, {Berta-Thompson}, {Law},
  {Schonhut-Stasik}, {Ziegler}, {Salama}, {Jensen-Clem}, {Duev}, {Riddle},
  {Kulkarni}, {Winters}, \& {Irwin}}]{2020AJ....159..139L}
{Lamman}, C., {Baranec}, C., {Berta-Thompson}, Z.~K., {et~al.} 2020, \aj, 159,
  139

\bibitem[{{Lee} {et~al.}(2013){Lee}, {Han}, \& {Park}}]{2013A&A...549A...2L}
{Lee}, B.~C., {Han}, I., \& {Park}, M.~G. 2013, \aap, 549, A2

\bibitem[{{Lomb}(1976)}]{1976Ap&SS..39..447L}
{Lomb}, N.~R. 1976, \apss, 39, 447

\bibitem[{{Lovis} {et~al.}(2011){Lovis}, {Dumusque}, {Santos}, {Bouchy},
  {Mayor}, {Pepe}, {Queloz}, {S{\'e}gransan}, \& {Udry}}]{2011arXiv1107.5325L}
{Lovis}, C., {Dumusque}, X., {Santos}, N.~C., {et~al.} 2011, arXiv e-prints
  [\eprint[arXiv]{1107.5325}]

\bibitem[{{Lovis} \& {Pepe}(2007)}]{2007A&A...468.1115L}
{Lovis}, C. \& {Pepe}, F. 2007, \aap, 468, 1115

\bibitem[{{Luque} {et~al.}(2018){Luque}, {Nowak}, {Pall{\'e}}, {Kossakowski},
  {Trifonov}, {Zechmeister}, {B{\'e}jar}, {Cardona Guill{\'e}n}, {Tal-Or},
  {Hidalgo}, {Ribas}, {Reiners}, {Caballero}, {Amado}, {Quirrenbach},
  {Aceituno}, {Cort{\'e}s-Contreras}, {D{\'{\i}}ez-Alonso}, {Dreizler},
  {Guenther}, {Henning}, {Jeffers}, {Kaminski}, {K{\"u}rster}, {Lafarga},
  {Montes}, {Morales}, {Passegger}, {Schmitt}, \&
  {Schweitzer}}]{2018A&A...620A.171L}
{Luque}, R., {Nowak}, G., {Pall{\'e}}, E., {et~al.} 2018, \aap, 620, A171

\bibitem[{{Maldonado} {et~al.}(2017){Maldonado}, {Scandariato}, {Stelzer},
  {Biazzo}, {Lanza}, {Maggio}, {Micela}, {Gonz{\'a}lez-{\'A}lvarez}, {Affer},
  {Claudi}, {Cosentino}, {Damasso}, {Desidera}, {Gonz{\'a}lez Hern{\'a}ndez},
  {Gratton}, {Leto}, {Messina}, {Molinari}, {Pagano}, {Perger}, {Piotto},
  {Rebolo}, {Ribas}, {Sozzetti}, {Su{\'a}rez Mascare{\~n}o}, \& {Zanmar
  Sanchez}}]{2017A&A...598A..27M}
{Maldonado}, J., {Scandariato}, G., {Stelzer}, B., {et~al.} 2017, \aap, 598,
  A27

\bibitem[{{Marcy} {et~al.}(1998){Marcy}, {Butler}, {Vogt}, {Fischer}, \&
  {Lissauer}}]{1998ApJ...505L.147M}
{Marcy}, G.~W., {Butler}, R.~P., {Vogt}, S.~S., {Fischer}, D., \& {Lissauer},
  J.~J. 1998, \apjl, 505, L147

\bibitem[{{Mayor} {et~al.}(2003){Mayor}, {Pepe}, {Queloz}, {Bouchy},
  {Rupprecht}, {Lo Curto}, {Avila}, {Benz}, {Bertaux}, {Bonfils}, {Dall},
  {Dekker}, {Delabre}, {Eckert}, {Fleury}, {Gilliotte}, {Gojak}, {Guzman},
  {Kohler}, {Lizon}, {Longinotti}, {Lovis}, {Megevand}, {Pasquini}, {Reyes},
  {Sivan}, {Sosnowska}, {Soto}, {Udry}, {van Kesteren}, {Weber}, \&
  {Weilenmann}}]{2003Msngr.114...20M}
{Mayor}, M., {Pepe}, F., {Queloz}, D., {et~al.} 2003, The Messenger, 114, 20

\bibitem[{{Mel{\'e}ndez} {et~al.}(2017){Mel{\'e}ndez}, {Bedell}, {Bean},
  {Ram{\'\i}rez}, {Asplund}, {Dreizler}, {Yan}, {Shi}, {Lind}, {Ferraz-Mello},
  {Galarza}, {dos Santos}, {Spina}, {Maia}, {Alves-Brito}, {Monroe}, \&
  {Casagrande}}]{2017A&A...597A..34M}
{Mel{\'e}ndez}, J., {Bedell}, M., {Bean}, J.~L., {et~al.} 2017, \aap, 597, A34

\bibitem[{{Morales} {et~al.}(2019){Morales}, {Mustill}, {Ribas}, {Davies},
  {Reiners}, {Bauer}, {Kossakowski}, {Herrero}, {Rodr{\'{\i}}guez},
  {L{\'o}pez-Gonz{\'a}lez}, {Rodr{\'{\i}}guez-L{\'o}pez}, {B{\'e}jar},
  {Gonz{\'a}lez-Cuesta}, {Luque}, {Pall{\'e}}, {Perger}, {Baroch}, {Johansen},
  {Klahr}, {Mordasini}, {Anglada-Escud{\'e}}, {Caballero},
  {Cort{\'e}s-Contreras}, {Dreizler}, {Lafarga}, {Nagel}, {Passegger},
  {Reffert}, {Rosich}, {Schweitzer}, {Tal-Or}, {Trifonov}, {Zechmeister},
  {Quirrenbach}, {Amado}, {Guenther}, {Hagen}, {Henning}, {Jeffers},
  {Kaminski}, {K{\"u}rster}, {Montes}, {Seifert}, {Abell{\'a}n}, {Abril},
  {Aceituno}, {Aceituno}, {Alonso-Floriano}, {Ammler-von Eiff}, {Antona},
  {Arroyo-Torres}, {Azzaro}, {Barrado}, {Becerril-Jarque}, {Ben{\'{\i}}tez},
  {Berdi{\~n}as}, {Bergond}, {Brinkm{\"o}ller}, {del Burgo}, {Burn},
  {Calvo-Ortega}, {Cano}, {C{\'a}rdenas}, {Guill{\'e}n}, {Carro}, {Casal},
  {Casanova}, {Casasayas-Barris}, {Chaturvedi}, {Cifuentes}, {Claret},
  {Colom{\'e}}, {Czesla}, {D{\'{\i}}ez-Alonso}, {Dorda}, {Emsenhuber},
  {Fern{\'a}ndez}, {Fern{\'a}ndez-Mart{\'{\i}}n}, {Ferro}, {Fuhrmeister},
  {Galad{\'{\i}}-Enr{\'{\i}}quez}, {Cava}, {Vargas}, {Garcia-Piquer}, {Gesa},
  {Gonz{\'a}lez-{\'A}lvarez}, {Hern{\'a}ndez}, {Gonz{\'a}lez-Peinado},
  {Gu{\`a}rdia}, {Guijarro}, {de Guindos}, {Hatzes}, {Hauschildt}, {Hedrosa},
  {Hermelo}, {Arabi}, {Otero}, {Hintz}, {Holgado}, {Huber}, {Huke}, {Johnson},
  {de Juan}, {Kehr}, {Kemmer}, {Kim}, {Kl{\"u}ter}, {Klutsch}, {Labarga},
  {Labiche}, {Lalitha}, {Lamp{\'o}n}, {Lara}, {Launhardt}, {L{\'a}zaro},
  {Lizon}, {Llamas}, {Lodieu}, {L{\'o}pez del Fresno}, {Salas},
  {L{\'o}pez-Santiago}, {Madinabeitia}, {Mall}, {Mancini}, {Mandel}, {Marfil},
  {Molina}, {Mart{\'{\i}}n}, {Mart{\'{\i}}n-Fern{\'a}ndez},
  {Mart{\'{\i}}n-Ruiz}, {Mart{\'{\i}}nez-Rodr{\'{\i}}guez}, {Marvin},
  {Mirabet}, {Moya}, {Naranjo}, {Nelson}, {Nortmann}, {Nowak}, {Ofir},
  {Pascual}, {Pavlov}, {Pedraz}, {Medialdea}, {P{\'e}rez-Calpena}, {Perryman},
  {Rabaza}, {Ballesta}, {Rebolo}, {Redondo}, {Rix}, {Rodler}, {Trinidad},
  {Sabotta}, {Sadegi}, {Salz}, {S{\'a}nchez-Blanco}, {Carrasco},
  {S{\'a}nchez-L{\'o}pez}, {Sanz-Forcada}, {Sarkis}, {Sarmiento},
  {Sch{\"a}fer}, {Schlecker}, {Schmitt}, {Sch{\"o}fer}, {Solano}, {Sota},
  {Stahl}, {Stock}, {Stuber}, {St{\"u}rmer}, {Su{\'a}rez}, {Tabernero},
  {Tulloch}, {Veredas}, {Vico-Linares}, {Vilardell}, {Wagner}, {Winkler},
  {Wolthoff}, {Yan}, \& {Osorio}}]{2019Sci...365.1441M}
{Morales}, J.~C., {Mustill}, A.~J., {Ribas}, I., {et~al.} 2019, Science, 365,
  1441

\bibitem[{{Noyes} {et~al.}(1984){Noyes}, {Hartmann}, {Baliunas}, {Duncan}, \&
  {Vaughan}}]{1984ApJ...279..763N}
{Noyes}, R.~W., {Hartmann}, L.~W., {Baliunas}, S.~L., {Duncan}, D.~K., \&
  {Vaughan}, A.~H. 1984, \apj, 279, 763

\bibitem[{{Otegi} {et~al.}(2020){Otegi}, {Bouchy}, \&
  {Helled}}]{2020A&A...634A..43O}
{Otegi}, J.~F., {Bouchy}, F., \& {Helled}, R. 2020, \aap, 634, A43

\bibitem[{{Passegger} {et~al.}(2018){Passegger}, {Reiners}, {Jeffers},
  {Wende-von Berg}, {Sch{\"o}fer}, {Caballero}, {Schweitzer}, {Amado},
  {B{\'e}jar}, {Cort{\'e}s-Contreras}, {Hatzes}, {K{\"u}rster}, {Montes},
  {Pedraz}, {Quirrenbach}, {Ribas}, \& {Seifert}}]{2018A&A...615A...6P}
{Passegger}, V.~M., {Reiners}, A., {Jeffers}, S.~V., {et~al.} 2018, \aap, 615,
  A6

\bibitem[{{Pearson}(1895)}]{1895RSPS...58..240P}
{Pearson}, K. 1895, Proceedings of the Royal Society of London Series I, 58,
  240

\bibitem[{{Perger} {et~al.}(2017{\natexlab{a}}){Perger},
  {Garc{\'{\i}}a-Piquer}, {Ribas}, {Morales}, {Affer}, {Micela}, {Damasso},
  {Su{\'a}rez-Mascare{\~n}o}, {Gonz{\'a}lez-Hern{\'a}ndez}, {Rebolo},
  {Herrero}, {Rosich}, {Lafarga}, {Bignamini}, {Sozzetti}, {Claudi},
  {Cosentino}, {Molinari}, {Maldonado}, {Maggio}, {Lanza}, {Poretti}, {Pagano},
  {Desidera}, {Gratton}, {Piotto}, {Bonomo}, {Martinez Fiorenzano}, {Giacobbe},
  {Malavolta}, {Nascimbeni}, {Rainer}, \& {Scandariato}}]{2017A&A...598A..26P}
{Perger}, M., {Garc{\'{\i}}a-Piquer}, A., {Ribas}, I., {et~al.}
  2017{\natexlab{a}}, \aap, 598, A26

\bibitem[{{Perger} {et~al.}(2017{\natexlab{b}}){Perger}, {Ribas}, {Damasso},
  {Morales}, {Affer}, {Su{\'a}rez Mascare{\~n}o}, {Micela}, {Maldonado},
  {Gonz{\'a}lez Hern{\'a}ndez}, {Rebolo}, {Scandariato}, {Leto}, {Zanmar
  Sanchez}, {Benatti}, {Bignamini}, {Borsa}, {Carbognani}, {Claudi},
  {Desidera}, {Esposito}, {Lafarga}, {Martinez Fiorenzano}, {Herrero},
  {Molinari}, {Nascimbeni}, {Pagano}, {Pedani}, {Poretti}, {Rainer}, {Rosich},
  {Sozzetti}, \& {Toledo-Padr{\'o}n}}]{2017A&A...608A..63P}
{Perger}, M., {Ribas}, I., {Damasso}, M., {et~al.} 2017{\natexlab{b}}, \aap,
  608, A63

\bibitem[{{Perger} {et~al.}(2019){Perger}, {Scandariato}, {Ribas}, {Morales},
  {Affer}, {Azzaro}, {Amado}, {Anglada-Escud{\'e}}, {Baroch}, {Barrado},
  {Bauer}, {B{\'e}jar}, {Caballero}, {Cort{\'e}s-Contreras}, {Damasso},
  {Dreizler}, {Gonz{\'a}lez-Cuesta}, {Gonz{\'a}lez Hern{\'a}ndez}, {Guenther},
  {Henning}, {Herrero}, {Jeffers}, {Kaminski}, {K{\"u}rster}, {Lafarga},
  {Leto}, {L{\'o}pez-Gonz{\'a}lez}, {Maldonado}, {Micela}, {Montes},
  {Pinamonti}, {Quirrenbach}, {Rebolo}, {Reiners}, {Rodr{\'{\i}}guez},
  {Rodr{\'{\i}}guez-L{\'o}pez}, {Schmitt}, {Sozzetti}, {Su{\'a}rez
  Mascare{\~n}o}, {Toledo-Padr{\'o}n}, {Zanmar S{\'a}nchez}, {Zapatero Osorio},
  \& {Zechmeister}}]{2019A&A...624A.123P}
{Perger}, M., {Scandariato}, G., {Ribas}, I., {et~al.} 2019, \aap, 624, A123

\bibitem[{Perrakis {et~al.}(2014)Perrakis, Ntzoufras, \&
  Tsionas}]{perrakis2014use}
Perrakis, K., Ntzoufras, I., \& Tsionas, E.~G. 2014, Computational Statistics
  \& Data Analysis, 77, 54

\bibitem[{{Pinamonti} {et~al.}(2018){Pinamonti}, {Damasso}, {Marzari},
  {Sozzetti}, {Desidera}, {Maldonado}, {Scandariato}, {Affer}, {Lanza},
  {Bignamini}, {Bonomo}, {Borsa}, {Claudi}, {Cosentino}, {Giacobbe},
  {Gonz{\'a}lez-{\'A}lvarez}, {Gonz{\'a}lez Hern{\'a}ndez}, {Gratton}, {Leto},
  {Malavolta}, {Martinez Fiorenzano}, {Micela}, {Molinari}, {Pagano}, {Pedani},
  {Perger}, {Piotto}, {Rebolo}, {Ribas}, {Su{\'a}rez Mascare{\~n}o}, \&
  {Toledo-Padr{\'o}n}}]{2018A&A...617A.104P}
{Pinamonti}, M., {Damasso}, M., {Marzari}, F., {et~al.} 2018, \aap, 617, A104

\bibitem[{{Pinamonti} {et~al.}(2019){Pinamonti}, {Sozzetti}, {Giacobbe},
  {Damasso}, {Scandariato}, {Perger}, {Gonz{\'a}lez Hern{\'a}ndez}, {Lanza},
  {Maldonado}, {Micela}, {Su{\'a}rez Mascare{\~n}o}, {Toledo-Padr{\'o}n},
  {Affer}, {Benatti}, {Bignamini}, {Bonomo}, {Claudi}, {Cosentino}, {Desidera},
  {Maggio}, {Martinez Fiorenzano}, {Pagano}, {Piotto}, {Rainer}, {Rebolo}, \&
  {Ribas}}]{2019A&A...625A.126P}
{Pinamonti}, M., {Sozzetti}, A., {Giacobbe}, P., {et~al.} 2019, \aap, 625, A126

\bibitem[{{Pojmanski}(1997)}]{1997AcA....47..467P}
{Pojmanski}, G. 1997, \actaa, 47, 467

\bibitem[{{Pollacco} {et~al.}(2006){Pollacco}, {Skillen}, {Collier Cameron},
  {Christian}, {Hellier}, {Irwin}, {Lister}, {Street}, {West}, {Anderson},
  {Clarkson}, {Deeg}, {Enoch}, {Evans}, {Fitzsimmons}, {Haswell}, {Hodgkin},
  {Horne}, {Kane}, {Keenan}, {Maxted}, {Norton}, {Osborne}, {Parley}, {Ryans},
  {Smalley}, {Wheatley}, \& {Wilson}}]{2006PASP..118.1407P}
{Pollacco}, D.~L., {Skillen}, I., {Collier Cameron}, A., {et~al.} 2006, \pasp,
  118, 1407

\bibitem[{{Pollack} {et~al.}(1996){Pollack}, {Hubickyj}, {Bodenheimer},
  {Lissauer}, {Podolak}, \& {Greenzweig}}]{1996Icar..124...62P}
{Pollack}, J.~B., {Hubickyj}, O., {Bodenheimer}, P., {et~al.} 1996, \icarus,
  124, 62

\bibitem[{{Quirrenbach} {et~al.}(2018){Quirrenbach}, {Amado}, {Ribas},
  {Reiners}, {Caballero}, {Seifert}, {Aceituno}, {Azzaro}, {Baroch}, {Barrado},
  {Bauer}, {Becerril}, {B{\`e}jar}, {Ben{\'\i}tez}, {Brinkm{\"o}ller}, {Cardona
  Guill{\'e}n}, {Cifuentes}, {Colom{\'e}}, {Cort{\'e}s-Contreras}, {Czesla},
  {Dreizler}, {Fr{\"o}lich}, {Fuhrmeister}, {Galad{\'\i}-Enr{\'\i}quez},
  {Gonz{\'a}lez Hern{\'a}ndez}, {Gonz{\'a}lez Peinado}, {Guenther}, {de
  Guindos}, {Hagen}, {Hatzes}, {Hauschildt}, {Helmling}, {Henning}, {Herbort},
  {Hern{\'a}ndez Casta{\~n}o}, {Herrero}, {Hintz}, {Jeffers}, {Johnson}, {de
  Juan}, {Kaminski}, {Klahr}, {K{\"u}rster}, {Lafarga}, {Sairam}, {Lamp{\'o}n},
  {Lara}, {Launhardt}, {L{\'o}pez del Fresno}, {L{\'o}pez-Puertas}, {Luque},
  {Mandel}, {Marfil}, {Mart{\'\i}n}, {Mart{\'\i}n-Ruiz}, {Mathar}, {Montes},
  {Morales}, {Nagel}, {Nortmann}, {Nowak}, {Pall{\'e}}, {Passegger}, {Pavlov},
  {Pedraz}, {P{\'e}rez-Medialdea}, {Perger}, {Rebolo}, {Reffert},
  {Rodr{\'\i}guez}, {Rodr{\'\i}guez L{\'o}pez}, {Rosich}, {Sabotta}, {Sadegi},
  {Salz}, {S{\'a}nchez-L{\'o}pez}, {Sanz-Forcada}, {Sarkis}, {Sch{\"a}fer},
  {Schiller}, {Schmitt}, {Sch{\"o}fer}, {Schweitzer}, {Shulyak}, {Solano},
  {Stahl}, {Tala Pinto}, {Trifonov}, {Zapatero Osorio}, {Yan}, {Zechmeister},
  {Abell{\'a}n}, {Abril}, {Alonso-Floriano}, {Ammler-von Eiff},
  {Anglada-Escud{\'e}}, {Anwand-Heerwart}, {Arroyo-Torres}, {Berdi{\~n}as},
  {Bergondy}, {Bl{\"u}mcke}, {del Burgo}, {Cano}, {Carro}, {C{\'a}rdenas},
  {Casal}, {Claret}, {D{\'\i}ez-Alonso}, {Doellinger}, {Dorda}, {Feiz},
  {Fern{\'a}ndez}, {Ferro}, {Gaisn{\'e}}, {Gallardo}, {G{\'a}lvez-Ortiz},
  {Garc{\'\i}a-Piquer}, {Garc{\'\i}a-Vargas}, {Garrido}, {Gesa}, {G{\'o}mez
  Galera}, {Gonz{\'a}lez-{\'A}lvarez}, {Gonz{\'a}lez-Cuesta}, {Grohnert},
  {Gr{\"o}zinger}, {Gu{\`a}rdia}, {Guijarro}, {Hedrosa}, {Hermann}, {Hermelo},
  {Hern{\'a}ndez Arab{\'\i}}, {Hern{\'a}ndez Hernando}, {Hidalgo}, {Holgado},
  {Huber}, {Huber}, {Huke}, {Kehr}, {Kim}, {Klein}, {Kl{\"u}ter}, {Klutsch},
  {Labarga}, {Labiche}, {Lamert}, {Laun}, {L{\'a}zaro}, {Lemke}, {Lenzen},
  {Llamas}, {Lizon}, {Lodieu}, {L{\'o}pez Gonz{\'a}lez}, {L{\'o}pez-Morales},
  {L{\'o}pez Salas}, {L{\'o}pez-Santiago}, {Mag{\'a}n Madinabeitia}, {Mall},
  {Mancini}, {Mar{\'\i}n Molina}, {Mart{\'\i}nez-Rodr{\'\i}guez}, {Maroto
  Fern{\'a}ndez}, {Marvin}, {Mirabet}, {Moreno-Raya}, {Moya}, {Mundt},
  {Naranjo}, {Panduro}, {Pascual}, {P{\'e}rez-Calpena}, {Perryman}, {Pluto},
  {Ram{\'o}n}, {Redondo}, {Reinhart}, {Rhode}, {Rix}, {Rodler}, {Rohloff},
  {S{\'a}nchez-Blanco}, {S{\'a}nchez Carrasco}, {Sarmiento}, {Schmidt},
  {Storz}, {Strachan}, {St{\"u}rmer}, {Su{\'a}rez}, {Tabernero}, {Tal-Or},
  {Tulloch}, {Ulbrich}, {Veredas}, {Vico Linares}, {Vidal-Dasilva},
  {Vilardell}, {Wagner}, {Winkler}, {Wolthoff}, {Xu}, \&
  {Zhao}}]{2018SPIE10702E..0WQ}
{Quirrenbach}, A., {Amado}, P.~J., {Ribas}, I., {et~al.} 2018, in Society of
  Photo-Optical Instrumentation Engineers (SPIE) Conference Series, Vol. 10702,
  Ground-based and Airborne Instrumentation for Astronomy VII, ed. C.~J.
  {Evans}, L.~{Simard}, \& H.~{Takami}, 107020W

\bibitem[{{Rainer} {et~al.}(2020){Rainer}, {Borsa}, \&
  {Affer}}]{2020ExA....49...73R}
{Rainer}, M., {Borsa}, F., \& {Affer}, L. 2020, Experimental Astronomy, 49, 73

\bibitem[{{Rajpaul} {et~al.}(2015){Rajpaul}, {Aigrain}, {Osborne}, {Reece}, \&
  {Roberts}}]{2015MNRAS.452.2269R}
{Rajpaul}, V., {Aigrain}, S., {Osborne}, M.~A., {Reece}, S., \& {Roberts}, S.
  2015, \mnras, 452, 2269

\bibitem[{{Reiners}(2009)}]{2009A&A...498..853R}
{Reiners}, A. 2009, \aap, 498, 853

\bibitem[{{Reiners} {et~al.}(2018){Reiners}, {Ribas}, {Zechmeister},
  {Caballero}, {Trifonov}, {Dreizler}, {Morales}, {Tal-Or}, {Lafarga},
  {Quirrenbach}, {Amado}, {Kaminski}, {Jeffers}, {Aceituno}, {B{\'e}jar},
  {Gu{\`a}rdia}, {Guenther}, {Hagen}, {Montes}, {Passegger}, {Seifert},
  {Schweitzer}, {Cort{\'e}s-Contreras}, {Abril}, {Alonso-Floriano}, {Ammler-von
  Eiff}, {Antona}, {Anglada-Escud{\'e}}, {Anwand-Heerwart}, {Arroyo-Torres},
  {Azzaro}, {Baroch}, {Barrado}, {Bauer}, {Becerril}, {Ben{\'{\i}}tez},
  {Berdi{\~n}as}, {Bergond}, {Bl{\"u}mcke}, {Brinkm{\"o}ller}, {del Burgo},
  {Cano}, {C{\'a}rdenas V{\'a}zquez}, {Casal}, {Cifuentes}, {Claret},
  {Colom{\'e}}, {Czesla}, {D{\'{\i}}ez-Alonso}, {Feiz}, {Fern{\'a}ndez},
  {Ferro}, {Fuhrmeister}, {Galad{\'{\i}}-Enr{\'{\i}}quez}, {Garcia-Piquer},
  {Garc{\'{\i}}a Vargas}, {Gesa}, {G{\'o}mez Galera}, {Gonz{\'a}lez
  Hern{\'a}ndez}, {Gonz{\'a}lez-Peinado}, {Gr{\"o}zinger}, {Grohnert},
  {Guijarro}, {de Guindos}, {Guti{\'e}rrez-Soto}, {Hatzes}, {Hauschildt},
  {Hedrosa}, {Helmling}, {Henning}, {Hermelo}, {Hern{\'a}ndez Arab{\'{\i}}},
  {Hern{\'a}ndez Casta{\~n}o}, {Hern{\'a}ndez Hernando}, {Herrero}, {Huber},
  {Huke}, {Johnson}, {de Juan}, {Kim}, {Klein}, {Kl{\"u}ter}, {Klutsch},
  {K{\"u}rster}, {Labarga}, {Lamert}, {Lamp{\'o}n}, {Lara}, {Laun}, {Lemke},
  {Lenzen}, {Launhardt}, {L{\'o}pez del Fresno}, {L{\'o}pez-Gonz{\'a}lez},
  {L{\'o}pez-Puertas}, {L{\'o}pez Salas}, {L{\'o}pez-Santiago}, {Luque},
  {Mag{\'a}n Madinabeitia}, {Mall}, {Mancini}, {Mandel}, {Marfil},
  {Mar{\'{\i}}n Molina}, {Maroto Fern{\'a}ndez}, {Mart{\'{\i}}n},
  {Mart{\'{\i}}n-Ruiz}, {Marvin}, {Mathar}, {Mirabet}, {Moreno-Raya}, {Moya},
  {Mundt}, {Nagel}, {Naranjo}, {Nortmann}, {Nowak}, {Ofir}, {Oreiro},
  {Pall{\'e}}, {Panduro}, {Pascual}, {Pavlov}, {Pedraz}, {P{\'e}rez-Calpena},
  {P{\'e}rez Medialdea}, {Perger}, {Perryman}, {Pluto}, {Rabaza}, {Ram{\'o}n},
  {Rebolo}, {Redondo}, {Reffert}, {Reinhart}, {Rhode}, {Rix}, {Rodler},
  {Rodr{\'{\i}}guez}, {Rodr{\'{\i}}guez-L{\'o}pez}, {Rodr{\'{\i}}guez
  Trinidad}, {Rohloff}, {Rosich}, {Sadegi}, {S{\'a}nchez-Blanco}, {S{\'a}nchez
  Carrasco}, {S{\'a}nchez-L{\'o}pez}, {Sanz-Forcada}, {Sarkis}, {Sarmiento},
  {Sch{\"a}fer}, {Schmitt}, {Schiller}, {Sch{\"o}fer}, {Solano}, {Stahl},
  {Strachan}, {St{\"u}rmer}, {Su{\'a}rez}, {Tabernero}, {Tala}, {Tulloch},
  {Ulbrich}, {Veredas}, {Vico Linares}, {Vilardell}, {Wagner}, {Winkler},
  {Wolthoff}, {Xu}, {Yan}, \& {Zapatero Osorio}}]{2018A&A...609L...5R}
{Reiners}, A., {Ribas}, I., {Zechmeister}, M., {et~al.} 2018, \aap, 609, L5

\bibitem[{{Ribas} {et~al.}(2018){Ribas}, {Tuomi}, {Reiners}, {Butler},
  {Morales}, {Perger}, {Dreizler}, {Rodr{\'\i}guez-L{\'o}pez}, {Gonz{\'a}lez
  Hern{\'a}ndez}, {Rosich}, {Feng}, {Trifonov}, {Vogt}, {Caballero}, {Hatzes},
  {Herrero}, {Jeffers}, {Lafarga}, {Murgas}, {Nelson}, {Rodr{\'\i}guez},
  {Strachan}, {Tal-Or}, {Teske}, {Toledo-Padr{\'o}n}, {Zechmeister},
  {Quirrenbach}, {Amado}, {Azzaro}, {B{\'e}jar}, {Barnes}, {Berdi{\~n}as},
  {Burt}, {Coleman}, {Cort{\'e}s-Contreras}, {Crane}, {Engle}, {Guinan},
  {Haswell}, {Henning}, {Holden}, {Jenkins}, {Jones}, {Kaminski}, {Kiraga},
  {K{\"u}rster}, {Lee}, {L{\'o}pez-Gonz{\'a}lez}, {Montes}, {Morin}, {Ofir},
  {Pall{\'e}}, {Rebolo}, {Reffert}, {Schweitzer}, {Seifert}, {Shectman},
  {Staab}, {Street}, {Su{\'a}rez Mascare{\~n}o}, {Tsapras}, {Wang}, \&
  {Anglada-Escud{\'e}}}]{2018Natur.563..365R}
{Ribas}, I., {Tuomi}, M., {Reiners}, A., {et~al.} 2018, \nat, 563, 365

\bibitem[{{Robertson} {et~al.}(2013{\natexlab{a}}){Robertson}, {Endl},
  {Cochran}, \& {Dodson-Robinson}}]{2013ApJ...764....3R}
{Robertson}, P., {Endl}, M., {Cochran}, W.~D., \& {Dodson-Robinson}, S.~E.
  2013{\natexlab{a}}, \apj, 764, 3

\bibitem[{{Robertson} {et~al.}(2013{\natexlab{b}}){Robertson}, {Endl},
  {Cochran}, {MacQueen}, \& {Boss}}]{2013ApJ...774..147R}
{Robertson}, P., {Endl}, M., {Cochran}, W.~D., {MacQueen}, P.~J., \& {Boss},
  A.~P. 2013{\natexlab{b}}, \apj, 774, 147

\bibitem[{{Robertson} {et~al.}(2014){Robertson}, {Mahadevan}, {Endl}, \&
  {Roy}}]{2014Sci...345..440R}
{Robertson}, P., {Mahadevan}, S., {Endl}, M., \& {Roy}, A. 2014, Science, 345,
  440

\bibitem[{{Savanov}(2012)}]{2012ARep...56..716S}
{Savanov}, I.~S. 2012, Astronomy Reports, 56, 716

\bibitem[{{Scandariato} {et~al.}(2017){Scandariato}, {Maldonado}, {Affer},
  {Biazzo}, {Leto}, {Stelzer}, {Zanmar Sanchez}, {Claudi}, {Cosentino},
  {Damasso}, {Desidera}, {Gonz{\'a}lez {\'A}lvarez}, {Gonz{\'a}lez
  Hern{\'a}ndez}, {Gratton}, {Lanza}, {Maggio}, {Messina}, {Micela}, {Pagano},
  {Perger}, {Piotto}, {Rebolo}, {Ribas}, {Rosich}, {Sozzetti}, \& {Su{\'a}rez
  Mascare{\~n}o}}]{2017A&A...598A..28S}
{Scandariato}, G., {Maldonado}, J., {Affer}, L., {et~al.} 2017, \aap, 598, A28

\bibitem[{Schraudolph {et~al.}(2007)Schraudolph, Yu, \&
  G{\"u}nter}]{schraudolph2007stochastic}
Schraudolph, N.~N., Yu, J., \& G{\"u}nter, S. 2007, in Artificial intelligence
  and statistics, 436--443

\bibitem[{{Selsis} {et~al.}(2007){Selsis}, {Kasting}, {Levrard}, {Paillet},
  {Ribas}, \& {Delfosse}}]{2007A&A...476.1373S}
{Selsis}, F., {Kasting}, J.~F., {Levrard}, B., {et~al.} 2007, \aap, 476, 1373

\bibitem[{{Sozzetti} {et~al.}(2013){Sozzetti}, {Bernagozzi}, {Bertolini},
  {Calcidese}, {Carbognani}, {Cenadelli}, {Christille}, {Damasso}, {Giacobbe},
  {Lanteri}, {Lattanzi}, \& {Smart}}]{2013EPJWC..4703006S}
{Sozzetti}, A., {Bernagozzi}, A., {Bertolini}, E., {et~al.} 2013, in European
  Physical Journal Web of Conferences, Vol.~47, European Physical Journal Web
  of Conferences, 03006

\bibitem[{{Su{\'a}rez Mascare{\~n}o} {et~al.}(2020){Su{\'a}rez Mascare{\~n}o},
  {Faria}, {Figueira}, {Lovis}, {Damasso}, {Gonz{\'a}lez Hern{\'a}ndez},
  {Rebolo}, {Cristiani}, {Pepe}, {Santos}, {Zapatero Osorio}, {Adibekyan},
  {Hojjatpanah}, {Sozzetti}, {Murgas}, {Abreu}, {Affolter}, {Alibert},
  {Aliverti}, {Allart}, {Allende Prieto}, {Alves}, {Amate}, {Avila}, {Baldini},
  {Bandi}, {Barros}, {Bianco}, {Benz}, {Bouchy}, {Broeng}, {Cabral},
  {Calderone}, {Cirami}, {Coelho}, {Conconi}, {Coretti}, {Cumani}, {Cupani},
  {D'Odorico}, {Deiries}, {Delabre}, {Di Marcantonio}, {Dumusque},
  {Ehrenreich}, {Fragoso}, {Genolet}, {Genoni}, {G{\'e}nova Santos}, {Hughes},
  {Iwert}, {Kerber}, {Knusdstrup}, {Landoni}, {Lavie}, {Lillo-Box}, {Lizon},
  {Lo Curto}, {Maire}, {Manescau}, {Martins}, {M{\'e}gevand}, {Mehner},
  {Micela}, {Modigliani}, {Molaro}, {Monteiro}, {Monteiro}, {Moschetti},
  {Mueller}, {Nunes}, {Oggioni}, {Oliveira}, {Pall{\'e}}, {Pariani},
  {Pasquini}, {Poretti}, {Rasilla}, {Redaelli}, {Riva}, {Santana Tschudi},
  {Santin}, {Santos}, {Segovia}, {Sosnowska}, {Sousa}, {Span{\`o}}, {Tenegi},
  {Udry}, {Zanutta}, \& {Zerbi}}]{2020A&A...639A..77S}
{Su{\'a}rez Mascare{\~n}o}, A., {Faria}, J.~P., {Figueira}, P., {et~al.} 2020,
  \aap, 639, A77

\bibitem[{{Su{\'a}rez Mascare{\~n}o} {et~al.}(2017){Su{\'a}rez Mascare{\~n}o},
  {Gonz{\'a}lez Hern{\'a}ndez}, {Rebolo}, {Velasco}, {Toledo-Padr{\'o}n},
  {Affer}, {Perger}, {Micela}, {Ribas}, {Maldonado}, {Leto}, {Zanmar Sanchez},
  {Scand ariato}, {Damasso}, {Sozzetti}, {Esposito}, {Covino}, {Maggio},
  {Lanza}, {Desidera}, {Rosich}, {Bignamini}, {Claudi}, {Benatti}, {Borsa},
  {Pedani}, {Molinari}, {Morales}, {Herrero}, \&
  {Lafarga}}]{2017A&A...605A..92S}
{Su{\'a}rez Mascare{\~n}o}, A., {Gonz{\'a}lez Hern{\'a}ndez}, J.~I., {Rebolo},
  R., {et~al.} 2017, \aap, 605, A92

\bibitem[{{Su{\'a}rez Mascare{\~n}o} {et~al.}(2016){Su{\'a}rez Mascare{\~n}o},
  {Rebolo}, \& {Gonz{\'a}lez Hern{\'a}ndez}}]{2016A&A...595A..12S}
{Su{\'a}rez Mascare{\~n}o}, A., {Rebolo}, R., \& {Gonz{\'a}lez Hern{\'a}ndez},
  J.~I. 2016, \aap, 595, A12

\bibitem[{{Su{\'a}rez Mascare{\~n}o} {et~al.}(2015){Su{\'a}rez Mascare{\~n}o},
  {Rebolo}, {Gonz{\'a}lez Hern{\'a}ndez}, \& {Esposito}}]{2015MNRAS.452.2745S}
{Su{\'a}rez Mascare{\~n}o}, A., {Rebolo}, R., {Gonz{\'a}lez Hern{\'a}ndez},
  J.~I., \& {Esposito}, M. 2015, \mnras, 452, 2745

\bibitem[{{Su{\'a}rez Mascare{\~n}o} {et~al.}(2018){Su{\'a}rez Mascare{\~n}o},
  {Rebolo}, {Gonz{\'a}lez Hern{\'a}ndez}, {Toledo-Padr{\'o}n}, {Perger},
  {Ribas}, {Affer}, {Micela}, {Damasso}, {Maldonado}, {Gonz{\'a}lez-Alvarez},
  {Leto}, {Pagano}, {Scandariato}, {Sozzetti}, {Lanza}, {Malavolta}, {Claudi},
  {Cosentino}, {Desidera}, {Giacobbe}, {Maggio}, {Rainer}, {Esposito},
  {Benatti}, {Pedani}, {Morales}, {Herrero}, {Lafarga}, {Rosich}, \&
  {Pinamonti}}]{2018A&A...612A..89S}
{Su{\'a}rez Mascare{\~n}o}, A., {Rebolo}, R., {Gonz{\'a}lez Hern{\'a}ndez},
  J.~I., {et~al.} 2018, \aap, 612, A89

\bibitem[{{Toledo-Padr{\'o}n} {et~al.}(2019){Toledo-Padr{\'o}n}, {Gonz{\'a}lez
  Hern{\'a}ndez}, {Rodr{\'{\i}}guez-L{\'o}pez}, {Su{\'a}rez Mascare{\~n}o},
  {Rebolo}, {Butler}, {Ribas}, {Anglada-Escud{\'e}}, {Johnson}, {Reiners},
  {Caballero}, {Quirrenbach}, {Amado}, {B{\'e}jar}, {Morales}, {Perger},
  {Jeffers}, {Vogt}, {Teske}, {Shectman}, {Crane}, {D{\'{\i}}az}, {Arriagada},
  {Holden}, {Burt}, {Rodr{\'{\i}}guez}, {Herrero}, {Murgas}, {Pall{\'e}},
  {Morales}, {L{\'o}pez-Gonz{\'a}lez}, {D{\'{\i}}ez Alonso}, {Tuomi}, {Kiraga},
  {Engle}, {Guinan}, {Strachan}, {Aceituno}, {Aceituno}, {Casanova},
  {Mart{\'{\i}}n-Ruiz}, {Montes}, {Ortiz}, {Sota}, {Briol}, {Barbieri},
  {Cervini}, {Deldem}, {Dubois}, {Hambsch}, {Harris}, {Kotnik}, {Logie},
  {Lopez}, {McNeely}, {Ogmen}, {P{\'e}rez}, {Rau}, {Rodr{\'{\i}}guez},
  {Urquijo}, \& {Vanaverbeke}}]{2019MNRAS.488.5145T}
{Toledo-Padr{\'o}n}, B., {Gonz{\'a}lez Hern{\'a}ndez}, J.~I.,
  {Rodr{\'{\i}}guez-L{\'o}pez}, C., {et~al.} 2019, \mnras, 488, 5145

\bibitem[{{Tuomi} {et~al.}(2014){Tuomi}, {Jones}, {Barnes},
  {Anglada-Escud{\'e}}, \& {Jenkins}}]{2014MNRAS.441.1545T}
{Tuomi}, M., {Jones}, H. R.~A., {Barnes}, J.~R., {Anglada-Escud{\'e}}, G., \&
  {Jenkins}, J.~S. 2014, \mnras, 441, 1545

\bibitem[{{Weiss} {et~al.}(2018){Weiss}, {Isaacson}, {Marcy}, {Howard},
  {Petigura}, {Fulton}, {Winn}, {Hirsch}, {Sinukoff}, {Rowe}, \& {California
  Kepler Survey}}]{2018AJ....156..254W}
{Weiss}, L.~M., {Isaacson}, H.~T., {Marcy}, G.~W., {et~al.} 2018, \aj, 156, 254

\bibitem[{{Weiss} \& {Marcy}(2014)}]{2014ApJ...783L...6W}
{Weiss}, L.~M. \& {Marcy}, G.~W. 2014, \apjl, 783, L6

\bibitem[{{Wildi} {et~al.}(2010){Wildi}, {Pepe}, {Chazelas}, {Lo Curto}, \&
  {Lovis}}]{2010SPIE.7735E..4XW}
{Wildi}, F., {Pepe}, F., {Chazelas}, B., {Lo Curto}, G., \& {Lovis}, C. 2010,
  in \procspie, Vol. 7735, Ground-based and Airborne Instrumentation for
  Astronomy III, 77354X

\bibitem[{{Winters} {et~al.}(2015){Winters}, {Henry}, {Lurie}, {Hambly}, {Jao},
  {Bartlett}, {Boyd}, {Dieterich}, {Finch}, {Hosey}, {Ianna}, {Riedel},
  {Slatten}, \& {Subasavage}}]{2015AJ....149....5W}
{Winters}, J.~G., {Henry}, T.~J., {Lurie}, J.~C., {et~al.} 2015, \aj, 149, 5

\bibitem[{{Zechmeister} {et~al.}(2019){Zechmeister}, {Dreizler}, {Ribas},
  {Reiners}, {Caballero}, {Bauer}, {B{\'e}jar}, {Gonz{\'a}lez-Cuesta},
  {Herrero}, {Lalitha}, \& et~al.}]{2019A&A...627A..49Z}
{Zechmeister}, M., {Dreizler}, S., {Ribas}, I., {et~al.} 2019, \aap, 627, A49

\bibitem[{{Zechmeister} \& {K{\"u}rster}(2009)}]{2009A&A...496..577Z}
{Zechmeister}, M. \& {K{\"u}rster}, M. 2009, \aap, 496, 577

\bibitem[{{Zechmeister} {et~al.}(2009){Zechmeister}, {K{\"u}rster}, \&
  {Endl}}]{2009A&A...505..859Z}
{Zechmeister}, M., {K{\"u}rster}, M., \& {Endl}, M. 2009, \aap, 505, 859

\bibitem[{{Zechmeister} {et~al.}(2018){Zechmeister}, {Reiners}, {Amado},
  {Azzaro}, {Bauer}, {B{\'e}jar}, {Caballero}, {Guenther}, {Hagen}, {Jeffers},
  {Kaminski}, {K{\"u}rster}, {Launhardt}, {Montes}, {Morales}, {Quirrenbach},
  {Reffert}, {Ribas}, {Seifert}, {Tal-Or}, \& {Wolthoff}}]{2018A&A...609A..12Z}
{Zechmeister}, M., {Reiners}, A., {Amado}, P.~J., {et~al.} 2018, \aap, 609, A12

\end{thebibliography}
%

\begin{appendix} 

\section{Additional figures}

\begin{table*}
\centering
        \caption{Priors and parameters related to the long-term and rotation signals obtained from the final RV MCMC analysis.}
        \label{tab:CycleRot_Properties}
        \begin{tabular}{lcc}
                \hline
                Parameter & Priors & Value  \\
                \hline
                \multicolumn{3}{c}{Cycle} \\
                \hline
                $K_{\rm cycle}$ [m\,s$^{-1}$] & $\mathcal{U}$ (0.01,     20.0) & 4.22$^{+0.90}_{-0.90}$  \\
                $P_{\rm cycle}$ [d]           & $\mathcal{U}$ (2300.0, 3900.0) & 3363$^{+230}_{-217}$ \\
                $T$ [d]                       & $\mathcal{U}$ (2500.0, 3600.0) & 261$^{+181}_{-161}$ \\
                \hline
                \multicolumn{3}{c}{Rotation} \\
                \hline
                $K_{\rm rot}^{2}$ & $\mathcal{LU}$ (8.0,  40.0) & 4.3$^{+1.8}_{-1.6}$\,m\,s$^{-1}$ $^{(*)}$ \\
                $P_{\rm rot}$ [d] & Fixed                       & 36.1$^{+1.9}_{-0.7}$                      \\
                $t_{\rm s}$ [d]   & $\mathcal{LU}$ (1.0, 300.0) & 5.2$^{+2.1}_{-1.4}$                       \\
                $\log ($C$)$      & $\mathcal{LU}$ (0.0,   1.0) & $-$21$^{+14}_{-14}$                       \\
                \hline
                \multicolumn{3}{c}{Remaining Parameters} \\
                \hline
                jitter$_{\rm HARPS}$ [m\,s$^{-1}$]    & $\mathcal{LU}$ (0.01,   4.0)  & 0.20$^{+0.85}_{-0.17}$ \\
                jitter$_{\rm HARPS-N}$ [m\,s$^{-1}$]  & $\mathcal{LU}$ (0.01,   4.0)  & 1.21$^{+0.53}_{-0.48}$ \\
                jitter$_{\rm CARMENES}$ [m\,s$^{-1}$] & $\mathcal{LU}$ (0.01,   4.0)  & 0.87$^{+0.59}_{-0.28}$ \\
                offset$_{\rm HARPS}$ [m\,s$^{-1}$]    & $\mathcal{U}$ ($-$15.0, 15.0) & $-$5.7$^{+1.1}_{-1.1}$ \\
                offset$_{\rm HARPS-N}$ [m\,s$^{-1}$]  & $\mathcal{U}$ ($-$15.0, 15.0) & 2.2$^{+1.1}_{-1.0}$    \\
                offset$_{\rm CARMENES}$ [m\,s$^{-1}$] & $\mathcal{U}$ ($-$15.0, 15.0) & $-$1.7$^{+1.7}_{-1.7}$ \\
                \hline
        \end{tabular}
        \begin{minipage}{17.5 cm}
{\footnotesize $^{(*)}$ This value was calculated as the root square of the $K_{\rm rot}^{2}$ posterior distribution results.}
\end{minipage}  
\end{table*}

\begin{figure*}
    \centering
        \includegraphics[width=18.5cm,page=1]{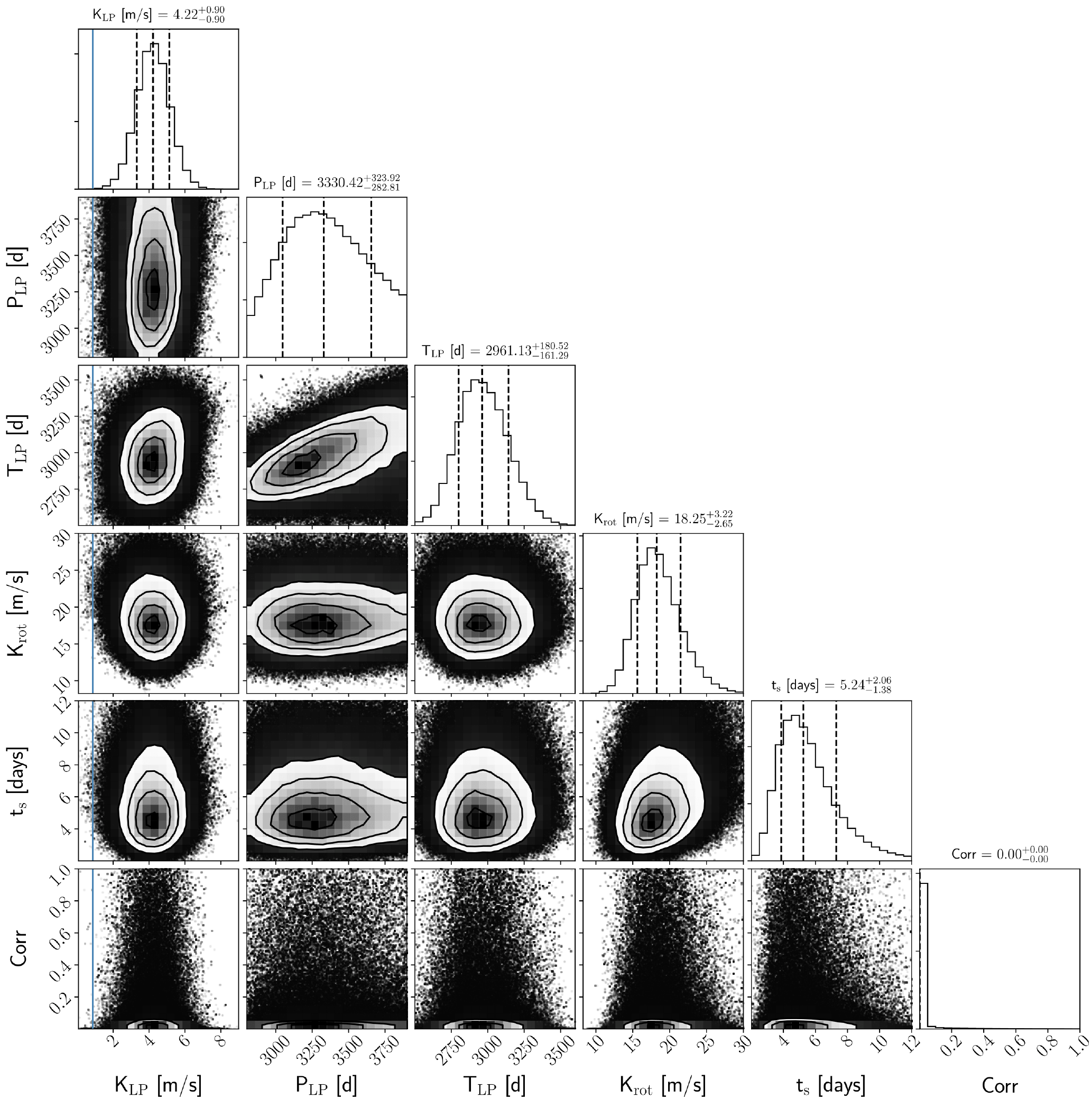} 
    \caption{Posterior distributions of the cycle and rotation fitted parameters from the cycle+rotation+Keplerian model applied to the RV time series. The 16th-84th percentiles are represented through vertical dashed lines.}
    \label{fig:RV_MCMC_SineRot1Pkepler_fixRot_3}
\end{figure*}

\begin{figure*}
        \includegraphics[width=18.5cm,page=1]{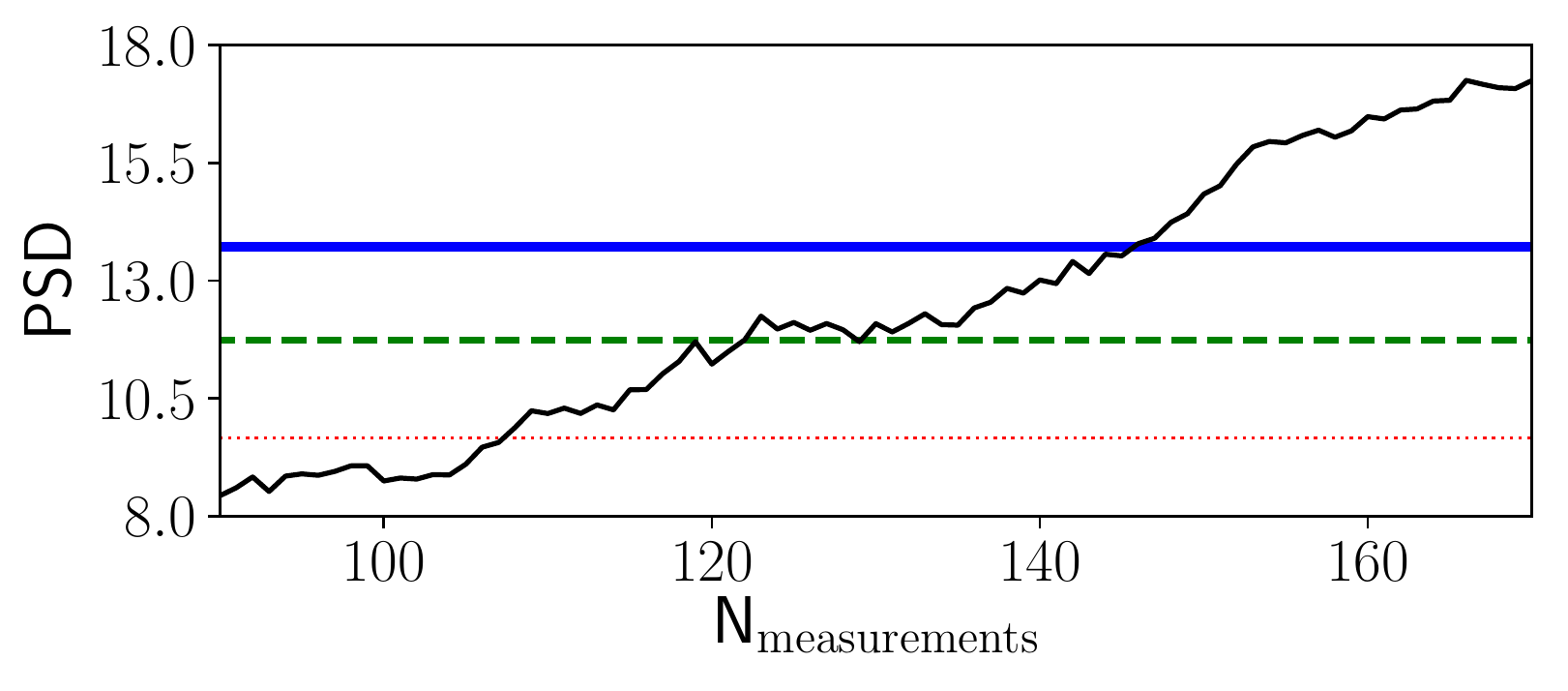}
    \caption{Evolution of the PSD associated with the planetary signal along the number of RV measurements considered. The blue, green, and red horizontal lines indicate the 0.1\%, 1\%, and 10\% FAP levels, respectively.}
    \label{fig:FAP_Evolution}
\end{figure*}

\end{appendix} 

\end{document}